\documentclass[a4paper]{article}

\usepackage[english]{babel}
\usepackage[utf8x]{inputenc}
\usepackage[T1]{fontenc}
\usepackage{booktabs}
\usepackage{subfig}
\usepackage[titletoc,title]{appendix}
\usepackage{amsfonts}
\usepackage{amssymb}
\usepackage{float}
\usepackage{array}
\usepackage{multirow}

\usepackage[a4paper,top=3cm,bottom=2cm,left=3cm,right=3cm,marginparwidth=1.75cm]{geometry}

\usepackage{amsmath}
\usepackage{graphicx}
\graphicspath{{figs/}}
\usepackage{authblk}

\title{fPINNs: Fractional Physics-Informed Neural Networks}

\date{}

\author{Guofei Pang$^*$}
\author{Lu Lu\thanks{The first two authors contributed equally to the work.}}
\author{George Em Karniadakis\thanks{Corresponding author: george\_karniadakis@brown.edu}}
\affil{Division of Applied Mathematics, Brown University, Providence, RI 02912, USA}

\begin{document}
\maketitle

\begin{abstract}
Physics-informed neural networks (PINNs), introduced in~\cite{raissi2018physics}, are effective in solving integer-order partial differential equations (PDEs) based on scattered and noisy data. PINNs employ standard feedforward neural networks (NNs) with the PDEs explicitly encoded into the NN using automatic differentiation, while the sum of the mean-squared PDE-residuals and the mean-squared error in initial/boundary conditions is minimized with respect to the NN parameters. Here we extend PINNs to fractional PINNs (fPINNs) to solve space-time fractional advection-diffusion equations (fractional ADEs), and we study systematically their convergence, hence explaining both of fPINNs and PINNs for first time. Specifically, we demonstrate their accuracy and effectiveness in solving multi-dimensional forward and inverse problems with forcing terms whose values are only known at randomly scattered spatio-temporal coordinates (\textit{black-box} forcing terms). A novel element of the fPINNs is the hybrid approach that we introduce for constructing the residual in the loss function using both automatic differentiation for the integer-order operators and numerical discretization for the fractional operators. This approach bypasses the difficulties stemming from the fact that automatic differentiation is not applicable to fractional operators because the standard chain rule in integer calculus is not valid in fractional calculus. To discretize the fractional operators, we employ the Gr\"unwald-Letnikov (GL) formula in one-dimensional fractional ADEs and the vector GL formula in conjunction with the directional fractional Laplacian in two- and three-dimensional fractional ADEs. We first consider the one-dimensional fractional Poisson equation and compare the convergence of the fPINNs against the finite difference method (FDM). We present the solution convergence using both the mean $L^2$ error as well as the standard deviation due to sensitivity to NN parameter initializations. Using different GL formulas we observe first-, second-, and third-order convergence rates for small size of training sets but the error saturates for larger training sets. We explain these results by analyzing the four sources of numerical errors due to discretization, sampling, NN approximation, and optimization. The total error decays monotonically (below $10^{-5}$ for third order GL formula) but it saturates beyond that point due to the optimization error. We also analyze the relative balance between discretization and sampling errors and observe that the sampling size and the number of discretization points (\textit{auxiliary points}) should be comparable to achieve the highest accuracy. As we increase the depth of the NN up to certain value, the mean error decreases and the standard deviation increases whereas the width has essentially no effect unless its value is either too small or too large. We next consider time-dependent fractional ADEs and compare white-box (WB) and black-box (BB) forcing. We observe that for the WB forcing, our results are similar to the aforementioned cases, however, for the BB forcing fPINNs outperform FDM. Subsequently, we consider multi-dimensional time-, space-, and space-time-fractional ADEs using the directional fractional Laplacian and we observe relative errors of $10^{-3}\sim10^{-4}$. Finally, we solve several inverse problems in 1D, 2D, and 3D to identify the fractional orders, diffusion coefficients, and transport velocities and obtain accurate results given proper initializations even in the presence of significant noise.


\end{abstract}

\section{Introduction}
There have been several applications of fractional partial differential equations in modeling anomalous transport, i.e., systems exhibiting memory effects, spatial nonlocality, or power-law characteristics. Among these are solute transport in fractured and porous media~\cite{benson2000application,sun2014use}, acoustic wave propagation with frequency-dependent dissipation~\cite{chen2004fractional,treeby2010modeling,holm2014comparison,zhu2014modeling}, laminar and turbulent flows~\cite{song2016fractional,uchaikin2013fractional,chen2006speculative}, viscoelastic constitutive laws~\cite{mainardi2010fractional}, just to mention a few. Fractional PDEs are phenomenological and hence they include parameters that need to be estimated using experiment data. These parameters are the differentiation orders of the fractional derivatives (namely, fractional orders), which may determine the power-law asymptotic behavior of certain characteristic responses. For instance, for sub-diffusion in the porous media, the order of the time-fractional derivative determines the decaying rate of the breakthrough curve for long-term observations. Identifying the fractional orders is  challenging as solving the forward problems already requires high computational cost due to the convolution expressions of fractional derivatives and hence the full matrices involved. On the other hand, the field- or experimental-measurements are usually sparse in spatio-temporal domain and may be polluted by noise as well. Identifying the fractional orders from sparse and noisy data is an important issue that has not been addressed adequately in the past.

Machine learning methods are particularly effective in solving data-driven forward and inverse problems of PDEs that involve black-box (BB) initial-boundary conditions and forcing terms. The BB here refers to function values measured only at specific spatio-temporal coordinates without explicit knowledge of the function. There have been some works on applying machine learning methods to discover the form of the integer-order PDEs \cite{brunton2016discovering,rudy2017data,long2018pde,gonzalez1998identification} using dictionaries of terms of various lengths. In this paper, we focus on the identification of the parameters of fractional PDEs whose overall form is known but some coefficients and most importantly the operators themselves are not known. The works on applying machine learning to parameter identification problems can be mainly divided into two categories. The first category exploits only the information of observed data in the spatio-temporal domain, and employs surrogate models such as Gaussian process regression (GP regression) \cite{pang2017discovering}, stochastic collocation methods \cite{yan2015stochastic}, and feedforward neural networks (NNs) \cite{garcia2006using,gorbachenko2016neural} to approximate the mapping from the parameters to be identified to the numerical solutions of PDEs or their mismatch with observed data. The PDEs are numerically solved to obtain the training points, and hence the information from PDEs is used implicitly. In contrast, the second category utilizes the information from PDEs {\it explicitly} by involving the differential operators of PDEs directly in the cost function to be optimized. The parameters to be identified, which appear in the differential operators, can be optimized by minimizing the cost function with respect to these parameters. For example, in \cite{raissi2017machine}, the parameters to be identified enter the negative log-likelihood function of the GP regression in the form of some extra hyperparameters of the covariance functions.
Subsequently, the same authors in \cite{raissi2018physics} added the parameters to be identified to the loss function of the NN and optimized these parameters jointly with the NN weights and biases. Other examples include, but not limited to, \cite{zhang2018quantifying,raissi2018hiddenphysics,gulian2018machine}.

\begin{figure}[H]
\centering
\includegraphics[width=.7\textwidth]{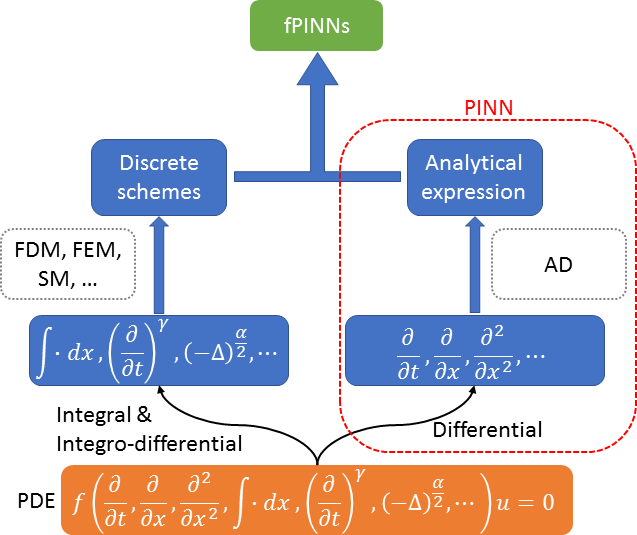}
\hfill
\caption{\label{Hybrid PINN}fPINNs for solving integral, differential, and integro-differential equations. Here we choose specific integro-differential operators in the form of time- and/or space- fractional derivatives. fPINNs can incorporate both fractional-order and integer-order operators. In the PDE shown in the figure, $f(\cdot)$ is a function of operators. The abbreviations ``SM'' and ``AD'' represent spectral methods and automatic differentiation, respectively.}
\end{figure}

In this paper, we focus on the NN approaches due to the high expressive power of NNs in function approximation~\cite{chen1993approximations,raghu2016expressive,lu2018collapse,pang2018neural}. In particular, we concentrate on physics-informed neural networks (PINNs)~\cite{dissanayake1994neural,van1995neural,lagaris1998artificial,raissi2018physics}, which belong to the second aforementioned category. The recent applications of PINNs include (1) inferring the velocity and pressure fields from the concentration field of a passive scalar in solving the Navier-Stokes equations~\cite{raissi2018hidden}, and (2) identifying the distributed parameters of stochastic PDEs~\cite{zhang2018quantifying}. However, PINNs, despite their high flexibility, cannot be directly applied to the solution of fractional PDEs, because the classical chain rule, which works rather efficiently in forward and backward propagation for NN, is not even valid in fractional calculus. We could consider a fractional version of chain rule, but it is in the form of an infinite series, and hence it is computationally prohibitive. To overcome this difficulty here we propose an alternative method in the form of fractional PINNs (fPINNs). Specifically, we propose fPINNs for solving integral, differential, and integro-differential equations, and more generally fPINNs can handle both fractional-order and integer-order operators. We employ the automatic differentiation technique to analytically derive the integer-order derivatives of NN output, while we approximate the fractional derivatives numerically using standard methods for the numerical discretization of fractional operators; an illustrative schematic is shown in Fig.~\ref{Hybrid PINN}. There are three attractive features of fPINNs.
\begin{itemize}
    \item[(1)]{\textbf{They have superior accuracy for black-box and noisy forcing terms.} When the forcing term is simply measured at scattered spatio-temporal points, interpolation has to be performed using standard numerical methods but this may introduce large interpolation errors for sparse measurements. In contrast, fPINNs can bypass the forcing term interpolation and instead construct the equation residual at these measurement points. Numerical results show that fPINNs can achieve higher solution accuracy for sparse measurements for both forward and inverse problems. Additionally, the noise in the data can be naturally taken into account by employing regularization techniques, such as $L^1$, $L^2$ and $L^{\infty}$ regularization~\cite{bengio2015deep}, early stopping~\cite{prechelt1998automatic}, as well as dropout~\cite{hinton2012improving,srivastava2014dropout}.}
    \item[(2)]{\textbf{They can easily handle high-dimensional, irregular-domain problems.} Being inherently data-driven, fPINNs do not rely on fixed meshes or grids, and thus they have higher flexibility in tackling high-dimensional problems on complex-geometry domains. The training points for fPINNs can be arbitrarily distributed in the spatio-temporal domain. We distinguish two groups of points: the training points and the points that help to calculate the fractional derivatives at the training points, which we term as the ``auxiliary" points. We will explain the concept of auxiliary points in Section~\ref{sec_fdm}. }
    \item[(3)]{\textbf{Same code for solving forward and inverse problems.} Minimum changes are needed to transform the forward problem code to an inverse problem code. We just need to add the parameters to be identified in inverse problem to the list of parameters to be optimized in the forward problem without changing anything else \cite{raissi2018physics}.}
\end{itemize}

We demonstrate the effectiveness of fPINNs by solving 1D, 2D, and 3D fractional ADEs for forward and inverse problems. We consider the fractional derivatives with respect to temporal and spatial variables. Specifically, we solve space-, time-, and space-time- fractional ADEs. To the best of our knowledge, we make the first attempt to solve the forward and the inverse problems for space-time-fractional ADEs defined in complex-geometry domains. For forward problems, there have been some works on solving 3D space-fractional ADEs. In ~\cite{lu2002possible,wang2013fast,wang2014fast,zhao2018fast}, only the 3D extension of the Riesz space-fractional derivatives is considered, which differs from the directional fractional Laplacian of our interest; in~\cite{minden2018simple}, Riesz fractional Laplacian is first regularized by singularity subtraction and then approximated by using trapezoidal rule. Specifically, in~\cite{minden2018simple}, equispaced nodes are required in order to make fast Fourier transform available; however, the method may be difficult to extend to non-equispaced nodes which have to be considered for a BB forcing. For inverse problems, little is known about 3D space-time-fractional ADEs with a fractional Laplacian, despite some existing works on 1D and 2D problems~\cite{zhang2011inverse,jin2012inverse,miller2013coefficient,tatar2013uniqueness,rundell2018recovering,guerngar2018inverse}. We note that the authors in~\cite{gulian2018machine} employed GP regression to identify the parameters in a multi-dimensional fractional Laplacian diffusion problems, but the fractional Laplacian was defined on an unbounded domain, namely $\mathbb{R}^d$ with spatial dimension $d$. Here, we consider the much more complex problem of a fractional Laplacian defined on a bounded domain.

The paper is organized as follows. In Section 2, we define forward and inverse problems for fractional ADEs. In Section 3, we first introduce the standard PINNs and then propose fPINNs. In the same section, we also review the finite difference schemes for approximating the fractional derivatives. In Section 4, we demonstrate the effectiveness of fPINNs using numerical examples. We first show the convergence rates of fPINNs and subsequently analyze the convergence in terms of four sources of errors, showing the solution accuracy for forward and inverse problems, followed by a discussion on the influence of noise in the data. Finally, we conclude the paper in Section 5.

\section{Fractional advection-diffusion equations (ADEs)} \label{sec_FADE}
We consider the following fractional ADE defined on a bounded domain $\Omega$ assuming  zero boundary conditions:
\begin{equation}\label{Forward_equation}
\begin{split}
\frac{\partial ^\gamma u(\boldsymbol{x},t)}{\partial t^\gamma} & = -c (-\Delta)^{\alpha/2}u(\boldsymbol{x},t)- \boldsymbol{v}\cdot \nabla u(\boldsymbol{x},t) + f_{BB}(\boldsymbol{x},t), \quad \boldsymbol{x}\in \Omega \subset \mathbb{R}^D, t\in (0,T],\\
u(\boldsymbol{x},t) & = 0, \quad \boldsymbol{x}\in \partial \Omega, \\
u(\boldsymbol{x},0) & = g(\boldsymbol{x}), \quad \boldsymbol{x}\in \Omega.
\end{split}
\end{equation}
The solution $u(\boldsymbol{x},t)$ is also assumed to be identically zero in the exterior of $\Omega$. The left-hand-side of the above equation is a time-fractional derivative of order $\gamma$, which is defined in the Caputo sense~\cite{kilbas2006theory}:
\begin{equation}\label{caputo_der}
    \frac{\partial^{\gamma}u(\boldsymbol{x},t)}{\partial t^{\gamma}} = \frac{1}{\Gamma(1-\gamma)}\int_0^{t}(t-\tau)^{-\gamma}\frac{\partial u(\boldsymbol{x},\tau)}{\partial \tau}d\tau, \quad 0<\gamma\leq 1,
\end{equation}
where $\Gamma(\cdot)$ is the gamma function. As $\gamma\rightarrow 1$ the time-fractional derivative reduces to the first derivative. The first term on the right-hand-side is a fractional Laplacian, which is defined in the sense of directional derivatives~\cite{lischke2018fractional,pang2015space}:
\begin{equation}\label{frac_lap}
(-\Delta)^{\alpha/2}u(\boldsymbol{x},t) = \frac{\Gamma\left(\frac{1-\alpha}{2}\right)\Gamma\left(\frac{D+\alpha}{2}\right)}{2\pi^{\frac{D+1}{2}}}\int_{||\boldsymbol{\theta}||_2=1}D^{\alpha}_{\boldsymbol{\theta}}u(\boldsymbol{x},t)d\boldsymbol{\theta}, \quad \boldsymbol{\theta}\in \mathbb{R}^D,\quad 1<\alpha \le 2,
\end{equation}
where $||\cdot||_2$ is the $L^2$ norm of a vector. The symbol $D^{\alpha}_{\boldsymbol{\theta}}$ denotes the directional fractional differential operator, where $\boldsymbol{\theta}$ is the differentiation direction vector. A review of this operator and its discretization will be given in Section \ref{sec_fdm}. As $\alpha\rightarrow2$ the fractional Laplacian (\ref{frac_lap}) reduces to the standard Laplaican `$-\Delta$'. In Problem (\ref{Forward_equation}) $\boldsymbol{v}$ is the mean-flow velocity and $f_{BB}$ is the BB forcing term whose values are only known at scattered spatio-temporal coordinates. In considering applications in groundwater contaminant transport, the fractional orders $\gamma$ and $\alpha$ have been restricted to $(0,1)$ and $(1,2)$, respectively. Also, for simplicity we consider zero boundary conditions.

The {\em forward problem} is formulated as: Given the fractional orders $\alpha$ and $\gamma$, the diffusion coefficient $c$, the flow velocity $\boldsymbol{v}$, the BB forcing term $f_{BB}$, as well as the initial and boundary conditions, we solve Problem (\ref{Forward_equation}) for the concentration field $u(\boldsymbol{x},t)$. On the other hand, the {\em inverse problem} is defined as: Given the initial-boundary conditions, the BB forcing term $f_{BB}$, and additional concentration measurements at the final time $u(\boldsymbol{x},T) = h_{BB}(\boldsymbol{x})$, we solve Problem (\ref{Forward_equation}) for the fractional orders $\alpha$ and $\gamma$, the diffusion coefficient $c$, the flow velocity $\boldsymbol{v}$, and the concentration field $u(\boldsymbol{x},t)$. We could also consider scattered measurements in time, but here we investigate a more challenging case having only available data at the final time.

\section{Methodology}\label{sec_PINN}
\subsection{Physics-Informed Neural Networks (PINNs)}
\begin{figure}[H]
\centering
\includegraphics[width=.7\textwidth]{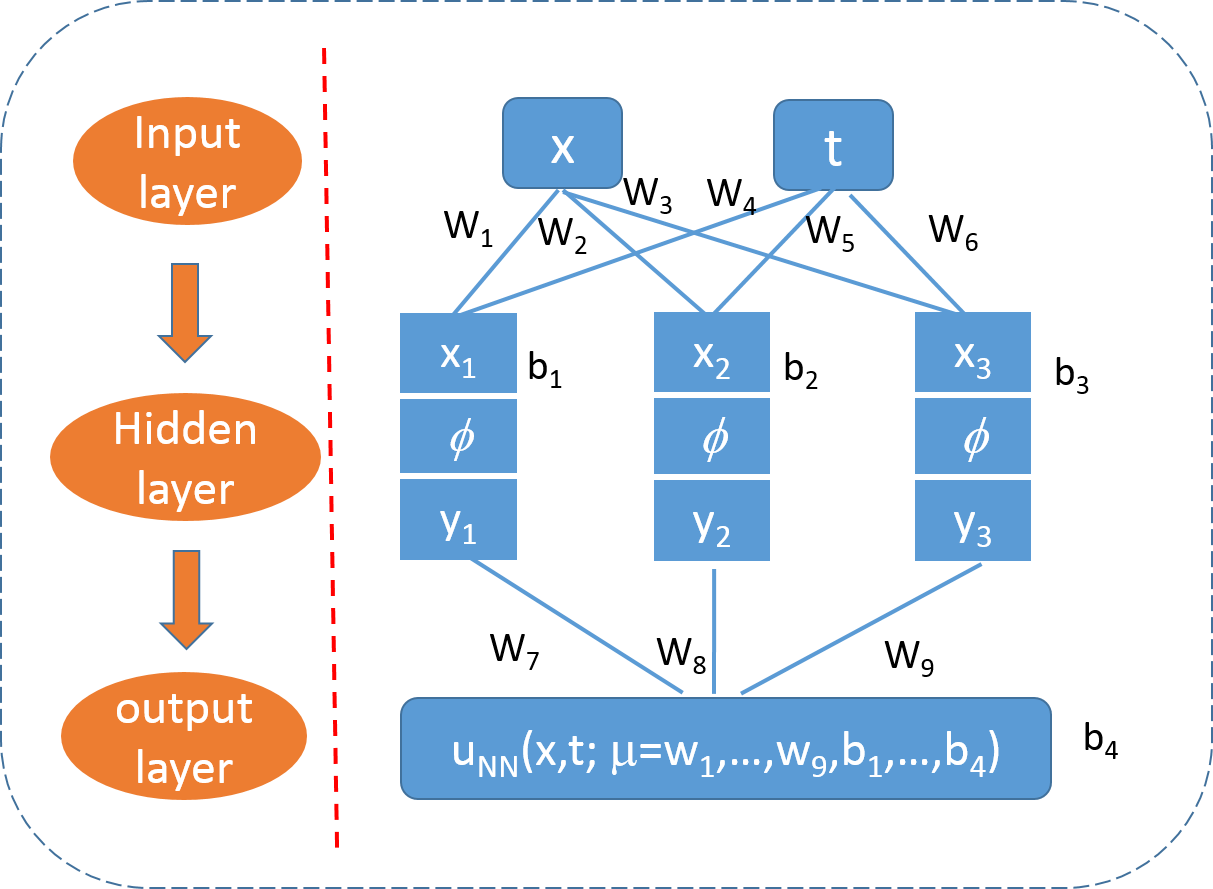}
\hfill
\caption{\label{NN-structure}\textit{Defining a simple NN $u_{NN}(x,t;\boldsymbol{\mu})$}. A fully connected NN with a single hidden layer consisting of three neurons. Each input of the hidden layer, $x_i$, is a linear transform of the inputs of the input layer: $x_1=W_1 x+W_4 t+b_1$, $x_2=W_2 x+W_5 t + b_2$, and $x_3=W_3 x+W_6 t +b_3$. Each output of the hidden layer, $y_i$,  is obtained after a nonlinear transform $\phi$ of the inputs: $y_i=\phi(x_i)$ for $i=1,2,3$. There are some candidate activation functions $\phi(\cdot)$ such as rectified linear unit (ReLU), sigmoid function, and hyperbolic function; in this paper, we adopt the third one. The final output is the linear transform of the outputs of the hidden layer: $u_{NN}(x,t)=W_7 y_1+W_8 y_2 + W_9 y_3+b_4$. The weights $W_i$'s and biases $b_i$'s constitute the parameter vector $\boldsymbol{\mu}$.    }
\end{figure}

To introduce the idea behind PINNs, we start with a 1D integer-order diffusion equation with zero boundary conditions:
\begin{equation}\label{1d-Poisson}
  \begin{split}
    \frac{\partial u(x,t)}{\partial t} & = \frac{\partial ^2 u(x,t)}{\partial x^2}+f_{BB}(x,t), \quad (x,t) \in (0,1)\times(0,1], \\
    u(x,0) & = g(x),\\
    u(0,t) & =  u(1,t) = 0.\\
  \end{split}
\end{equation}
Based on whether or not we enforce the initial-boundary conditions, there are three ways to construct the approximate solution $\tilde{u}(x,t)$ to the equation. In the first case, we can directly assume the approximate solution to be an output of NN, namely, $\tilde{u}(x,t)=u_{NN}(x,t;\boldsymbol{\mu})$. The NN here plays the role of surrogate model that approximates the mapping from the spatio-temporal coordinates to the solution of equation. The NN is parameterized by its weights and biases that constitute the parameter vector $\boldsymbol{\mu}$; see Fig.~\ref{NN-structure} for a simple NN with a single hidden layer. Note that in the numerical examples of the paper, we adopt NNs with multiple hidden layers. In the PINNs, we expect to optimize the NN parameters $\boldsymbol{\mu}$ such that the resulting approximate solution will satisfy both the equation and the initial-boundary conditions as close as possible~\cite{raissi2018physics}. In the second case, we can choose a form of the approximate solution that satisfies the boundary conditions automatically, namely, $\tilde{u}(x,t)=\rho(x)u_{NN}(x,t;\boldsymbol{\mu})$, where $\rho(0)=\rho(1)=0$ and the auxiliary function $\rho(x)$ is pre-selected. Provided that the initial condition function $g(x)$ is a WB function, the third case is to write the approximate solution as $\tilde{u}(x,t)=t\rho(x)u_{NN}(x,t;\boldsymbol{\mu})+g(x)$, where $\rho(0)=\rho(1)=0$. This approximate solution satisfies both initial and boundary conditions automatically. In this paper, we only focus on the third case, which yields a succinct form of loss function. It is straightforward to consider the other two cases.

\begin{figure}[H]
\centering
\includegraphics[width=.7\textwidth]{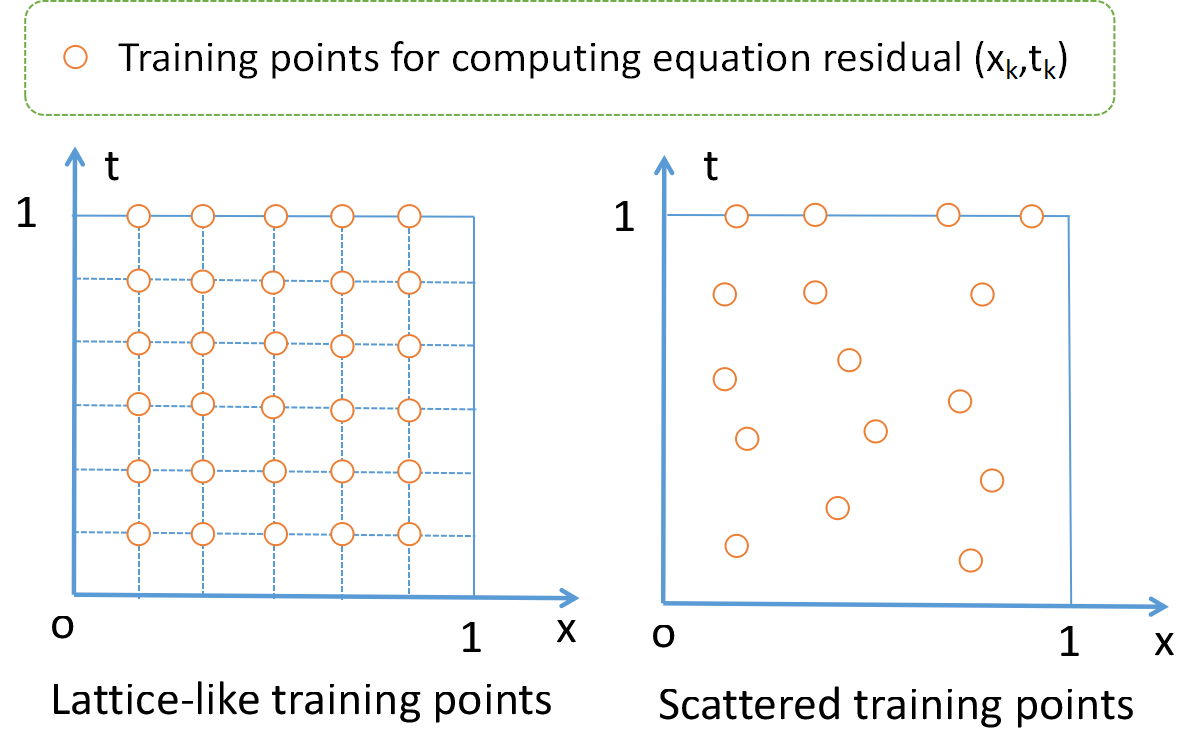}
\hfill
\caption{\label{training_pts1}\textit{Two distributions of training points of PINNs for 1D time-dependent diffusion equation.} Left: lattice-like training points coinciding with finite difference grid. Right: Scattered training points drawn from a quasi-random sequence.}
\end{figure}

The loss function of PINNs for the forward problem with the approximate solution
\begin{equation}
  \tilde{u}(x,t)=t\rho(x)u_{NN}(x,t;\boldsymbol{\mu})+g(x)
\end{equation}
  is defined as the mean-squared-error of the equation residual:
\begin{equation}\label{loss_fun}
     L(\boldsymbol{\mu})=\frac{1}{N}\sum_{k=1}^{N}\left(\frac{\partial \tilde{u}(x_k,t_k)}{\partial t}-\frac{\partial^2 \tilde{u}(x_k,t_k)}{\partial x^2}-f_{BB}(x_k,t_k)\right)^2.
\end{equation}
We use $N$ training points $\{(x_k,t_k)\},k=1,2,\cdots,N$ in the spatio-temporal domain to obtain the equation residual information. Each training point is an independent input vector $[x,t]^T$ of the NN plotted in Fig.~\ref{NN-structure}. We minimize the loss function with respect to $\boldsymbol{\mu}$ and we refer to this minimization procedure as ``training". The distribution of training points has certain impact on the flexibility of PINNs. Fig.~\ref{training_pts1} shows two different ways to select the training points. The lattice-like training points are exactly the same as the finite difference grid points, which are equispaced in the spatio-temporal domain. The scattered training points can be taken from certain quasi-random sequences, such as the Sobol sequences or the Latin hypercube sampling. The advantage of the scattered training points is that they provide more flexiblity for high-dimensional problems on irregular domains. Therefore, we will adopt the scattered training points in most examples of the paper except the cases where we compare fPINN to FDM.

We employ automatic differentiation in PINNs to compute the temporal and spatial derivatives of $\tilde{u}(x,t)$ in the loss function~(\ref{loss_fun}). To compute these derivatives, we have to evaluate the derivatives of $u_{NN}(x,t;\boldsymbol{\mu})$ first. We consider $\frac{\partial u_{NN}(x,t)}{\partial t}$ as an example, omitting $\boldsymbol{\mu}$ in $u_{NN}$ for simplicity of notation. From the caption of Fig. \ref{NN-structure}, we know that
\begin{equation}
    \begin{split}
        \frac{\partial u_{NN}(x,t)}{\partial t} & = W_7 \frac{\partial y_1(x,t)}{\partial t}+W_8 \frac{\partial y_2(x,t)}{\partial t}+W_9 \frac{\partial y_3(x,t)}{\partial t}\\
        \frac{\partial y_i(x,t)}{\partial t}&=\frac{\partial \phi(x_i(x,t))}{\partial x_i}\frac{\partial x_i(x,t)}{\partial t} =\frac{\partial \phi(x_i(x,t))}{\partial x_i}W_{k(i)},\quad i=1,2,3,
    \end{split}
\end{equation}
where $k(1)=4$, $k(2)=5$, and $k(3)=6$. The chain rule is used here to calculate $\partial y_i(x,t)/\partial t$. For multiple hidden layers, the chain rule is employed in the automatic differentiation to compute the derivatives hierarchically from the output layer to the input layer, which turns out to be rather efficient for very deep NNs. Nevertheless, there are some cases where the classical chain rule does not work. A typical example is the fractional derivatives of our interest, whose chain rule, if any, takes the form of infinite series~\cite{podlubny1999fractional,diethelm2010analysis,tarasov2016chain}; see Appendix A for comparison of chain rules for the integer-order and the fractional derivatives. Even worse, the backward propagation in computing the gradients of loss function with respect to weights and biases has to use the classical chain rule repeatedly for the integer-order derivatives, which for fractional derivatives is computationally prohibitive. To bypass the automatic differentiation for fractional derivatives, we replace the fractional differential operators with their discrete versions and then incorporate the discrete schemes into the loss function of PINNs. We refer to the resulting PINNs as fPINNs.

\subsection{Fractional PINNs (fPINNs)}

Before proceeding with the inverse problem, we first consider the forward problem of the form
\begin{equation}
  \begin{split}
    \mathcal{L}\{u(\boldsymbol{x},t)\} & = f_{BB}(\boldsymbol{x},t), \quad (\boldsymbol{x},t) \in \Omega \times (0,T],\\
    u(\boldsymbol{x},0) &= g(\boldsymbol{x}), \quad \boldsymbol{x} \in \Omega,\\
    u(\boldsymbol{x},t)&=0,\quad \boldsymbol{x}\in \partial \Omega,
  \end{split}
\end{equation}
where $g(\cdot)$ is assumed to be a WB initial condition function. The approximate solution is chosen as
\begin{equation}\label{approximate_solution}
    \tilde{u}(\boldsymbol{x},t)=t\rho(\boldsymbol{x})u_{NN}(\boldsymbol{x},t;\boldsymbol{\mu})+g(\boldsymbol{x}),
\end{equation}
such that it satisfies the initial-boundary conditions automatically. $\mathcal{L}\{\cdot\}$ is a linear or nonlinear operator. Here we consider $\mathcal{L}=\frac{\partial^{\gamma}}{\partial t^{\gamma}}+c(-\Delta)^{\alpha/2}+\boldsymbol{v}\cdot \nabla$, and we divide the component operators of $\mathcal{L}$ into two categories $\mathcal{L}=\mathcal{L}_{AD}+\mathcal{L}_{nonAD}$. The first category includes the operators that can be automatically differentiated (AD) using the classical chain rule. In other words, we have
\begin{equation}
    \mathcal{L}_{AD}:=\left\{\begin{array}{c}\boldsymbol{v}\cdot \nabla,\quad \alpha\in (1,2),\quad\gamma\in(0,1), \\ \frac{\partial}{\partial t}+\boldsymbol{v}\cdot \nabla,\quad \alpha\in (1,2),\quad \gamma=1,\\-\Delta+\boldsymbol{v}\cdot \nabla,\quad \alpha=2,\quad \gamma\in(0,1).\end{array}\right.
\end{equation}
The second category includes those operators that cannot be automatically differentiated, say, $\mathcal{L}_{nonAD}=\frac{\partial^{\gamma}}{\partial t^{\gamma}}+c(-\Delta)^{\alpha/2}$ for $\alpha\in(1,2), \gamma\in(0,1)$. For $\mathcal{L}_{nonAD}$, we can discretize it using standard numerical methods such as finite difference, finite element, or spectral methods, but here we focus on the FDM. We denote by $\mathcal{L}_{FDM}$ the FDM-discretization version of $\mathcal{L}_{nonAD}$ and then define the loss function $L_{FW}$ of fPINNs for the forward problem as
\begin{equation}\label{frac_PINN_loss}
    L_{FW}(\boldsymbol{\mu}) = \frac{1}{|\Xi|}\sum_{(\boldsymbol{x},t)\in \Xi}\left[\mathcal{L}_{FDM}\{\tilde{u}(\boldsymbol{x},t)\}+\mathcal{L}_{AD}\{\tilde{u}(\boldsymbol{x},t)\}-f_{BB}(\boldsymbol{x},t)\right]^2,
\end{equation}
where $|\Xi|$ represents the number of training points in the training set $\Xi\subset \Omega\times (0,T]$. We also note that if $\mathcal{L}_{nonAD}$ vanishes, fPINNs reduce to PINNs.
The training procedure of fPINNs for forward problems is to minimize the loss function with respect to $\boldsymbol{\mu}$ in order to obtain the best network parameters $\boldsymbol{\mu}_{opt}$. Then, we can use the trained NN to make predictions at arbitrary test points $(\boldsymbol{x}_{test},t_{test})$, namely, $\tilde{u}(\boldsymbol{x}_{test},t_{test};\boldsymbol{\mu}_{opt})$.

On the other hand, the inverse problem takes the form
\begin{equation}
  \begin{split}
    \mathcal{L}^{\boldsymbol{\xi}}\{u(\boldsymbol{x},t)\} & = f_{BB}(\boldsymbol{x},t), \quad (\boldsymbol{x},t) \in \Omega \times (0,T],\\
    u(\boldsymbol{x},0) &= g(\boldsymbol{x}), \quad \boldsymbol{x} \in \Omega,\\
    u(\boldsymbol{x},t) &=0, \quad \boldsymbol{x} \in \partial \Omega,\\
    u(\boldsymbol{x},t) &= h_{BB}(\boldsymbol{x},t), \quad (\boldsymbol{x},t) \in \Omega \times \{t=T\},
  \end{split}
\end{equation}
where the PDE parameters $\boldsymbol{\xi}$ and the concentration field $u(\cdot)$ are both to be recovered from the boundary and initial-final conditions. The loss function for the inverse problems is similar to that for the forward problems except that a final condition mismatch term is added and that the PDE parameters $\boldsymbol{\xi}$ are jointly optimized with the network parameters $\boldsymbol{\mu}$. Specifically, the loss function $L_{INV}$ for the inverse problem we consider is
\begin{equation}\label{frac_PINN_loss_inv}
  \begin{split}
    L_{INV}(\boldsymbol{\mu},\boldsymbol{\xi}=\{\alpha,\gamma,c,\boldsymbol{v}\}) &= w_1\cdot\frac{1}{|\Xi_1|}\sum_{(\boldsymbol{x},t)\in \Xi_1}\left[\mathcal{L}_{FDM}^{\alpha,\gamma,c}\{\tilde{u}(\boldsymbol{x},t)\}+\mathcal{L}_{AD}^{\boldsymbol{v}}\{\tilde{u}(\boldsymbol{x},t)\}-f_{BB}(\boldsymbol{x},t)\right]^2 \\
    & + w_2 \cdot \frac{1}{|\Xi_2|}\sum_{(\boldsymbol{x},t)\in\Xi_2}\left[\tilde{u}(\boldsymbol{x},t)-h_{BB}(\boldsymbol{x},t)\right]^2,
  \end{split}
\end{equation}
where in the current loss function we assume $\alpha\in(1,2)$ and $\gamma\in(0,1)$.
$\Xi_1\subset \Omega\times(0,T)$ and $\Xi_2\subset\Omega\times\{t=T\}$ are two distinct sets of training points, and $w_1$ and $w_2$ are pre-fixed weight factors that determine the relative contribution of each term. Optimizing the loss function yields the identified fractional orders $\alpha_{opt}$ and $\gamma_{opt}$, the diffusion coefficient $c_{opt}$, the flow velocity $\boldsymbol{v}_{opt}$, and the network parameters $\boldsymbol{\mu}_{opt}$. The concentration field is recovered to be $\tilde{u}(\boldsymbol{x}_{test},t_{test};\boldsymbol{\mu}_{opt})$.

\subsection{Finite difference schemes for fractional derivatives}\label{sec_fdm}
\begin{figure}[H]
\centering
\subfloat[1D problem]{
\includegraphics[width=.6\textwidth]{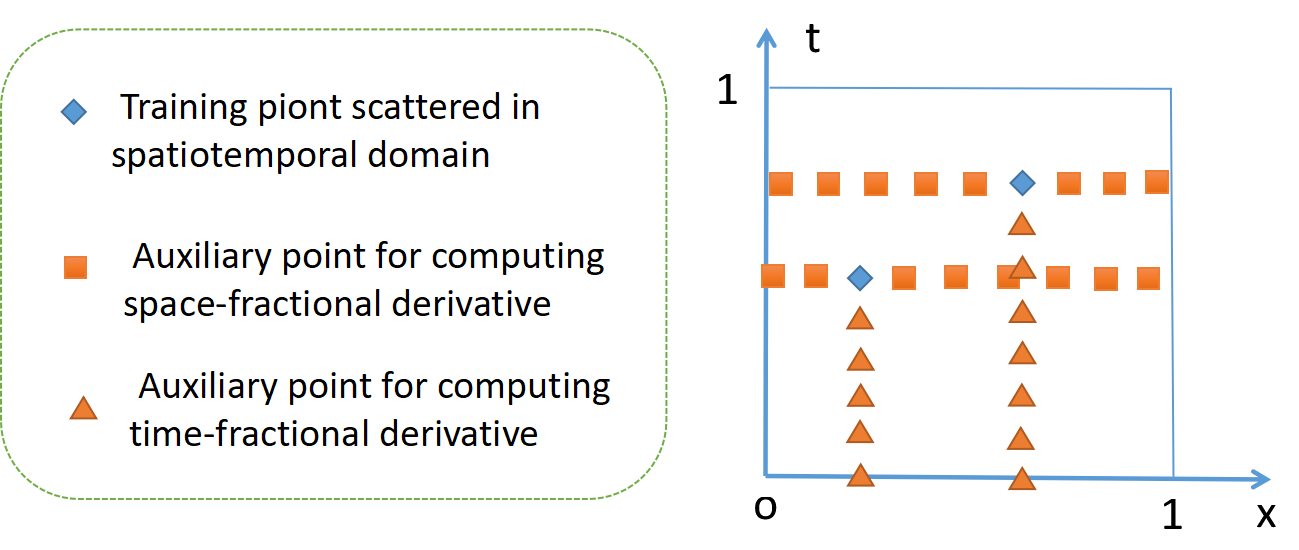}}\hfill
\subfloat[2D problem]{
\includegraphics[width=.38\textwidth]{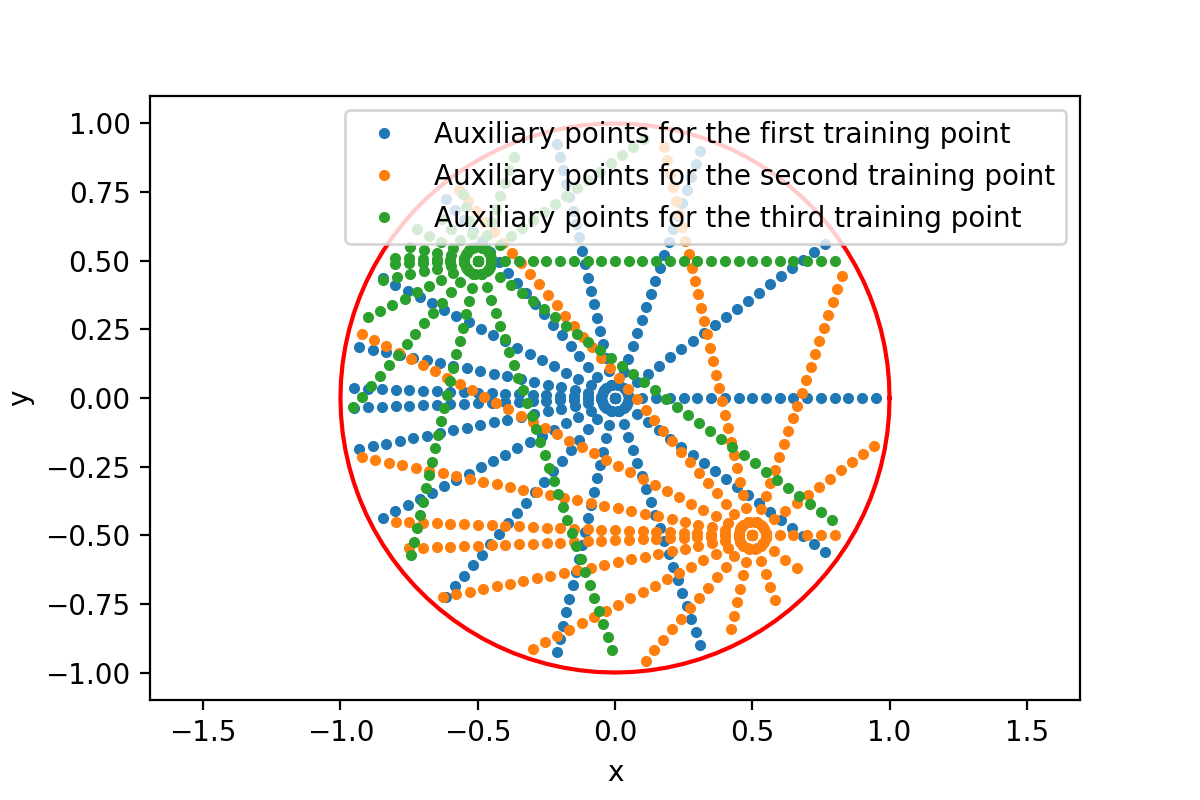}}\hfill
\caption{\label{training-auxiliary}\textit{Scattered training points and auxiliary points for fPINNs.} (a) Training and auxiliary points for the 1D space-time-fractional problem. (b) Auxiliary points for three different training points for 2D space-time-fractional problem defined on a unit disk (for fixed $t$). The training points are located at the centers of clusters of auxiliary points. A total of 15 Gauss-Legendre points are used in approximating $\int_0^{2\pi}(\cdot)d\theta$ in $(-\Delta)^{\alpha/2}$, and $\Delta x\approx 1/\lambda=0.05$.}
\end{figure}

For $\alpha\in(1,2)$ and $\gamma\in(0,1)$, the FDM-based discrete operator for the fractional ADE (\ref{Forward_equation}) is $\mathcal{L}_{FDM}=\mathcal{L}_{\Delta t}^{\gamma}+c\mathcal{L}_{\Delta x}^{\alpha}$. To approximate the time-fractional derivative in the equation, we adopt the commonly used finite difference $L_1$ scheme~\cite{sun2006fully,jin2015analysis}, as follows
\begin{multline}\label{Lt}
    \frac{\partial^{\gamma}\tilde{u}(\boldsymbol{x},t)}{\partial t^{\gamma}} \approx \mathcal{L}^{\gamma}_{\Delta t}\tilde{u}(\boldsymbol{x},t):=\frac{1}{\Gamma(2-\gamma)(\Delta t)^{\gamma}}\left\{-c_{\lceil \lambda t\rceil-1}\tilde{u}(\boldsymbol{x},0)+c_0 \tilde{u}(\boldsymbol{x},t) \right. \\
    \left. + \sum_{k=1}^{\lceil \lambda t\rceil-1}(c_{\lceil \lambda t\rceil-k}-c_{\lceil \lambda t\rceil-k-1})\tilde{u}(\boldsymbol{x},k\Delta t) \right\},
\end{multline}
where $c_l=(l+1)^{1-\gamma}-l^{1-\gamma}$. The temporal step is $\Delta t = t/\lceil \lambda t\rceil\approx 1/\lambda$, where $\lceil \cdot \rceil$ is the ceiling function, and the constant factor $\lambda$ determines the step size. The truncating error for $L_1$ scheme is $(\Delta t)^{2-\gamma}$~\cite{sun2006fully}. Given the spatial location $\boldsymbol{x}$, we see from the scheme that the time-fractional derivative of $\tilde{u}(\boldsymbol{x},t)$ evaluated at time $t$ depends on all the values of $\tilde{u}(\boldsymbol{x},t)$ evaluated at all the previous time steps $0,\Delta t,2\Delta t,\cdots,t$. We call the current time $t$ and the previous times $k\Delta t$ the training and auxiliary points, respectively. There are $\lceil\lambda t\rceil+1$ auxiliary points corresponding to the training point $t$. The use of $\lambda$ can allow scattered training points for both temporal and spatial discretizations. Fig.~\ref{training-auxiliary} (a) uses triangles to represent the auxiliary points for computing the time-fractional derivatives. If the equation includes also space-fractional derivatives, we need to consider the auxiliary points in spatial directions, which are shown by squares in Fig.~\ref{training-auxiliary}.

The finite difference scheme for the fractional Laplacian is more complicated than that for the time-fractional derivative. It is the key step to discretize the directional fractional derivative in the fractional Laplacian. The definition of the Riemann-Liouville directional derivative of a sufficiently properly defined function $w(\boldsymbol{x})$ is ($\alpha\in(1,2)$)~\cite{samko1993fractional,lischke2018fractional}
\begin{equation}
D_{\boldsymbol{\theta}}^{\alpha}w(\boldsymbol{x})=\frac{1}{\Gamma(2-\alpha)}\left(\boldsymbol{\theta}\cdot \nabla\right)^2\int_0^{+\infty}\xi^{1-\alpha}w(\boldsymbol{x}-\xi\boldsymbol{\theta})d\xi, \quad \boldsymbol{x}, \boldsymbol{\theta}\in \mathbb{R}^D,
\end{equation}
where the differentiation direction is defined as $\boldsymbol{\theta}=\cos\theta=\pm 1$ where $\theta=0$ or $\pi$ for the 1D case, $\boldsymbol{\theta}=\left[\cos\theta,\sin\theta\right],\theta\in \left[0,2\pi\right)$ for the 2D case, and $\boldsymbol{\theta}=\left[\sin\phi\cos\theta,\sin\phi\sin\theta,\cos\phi\right], \theta\in \left[0,2\pi\right), \phi\in\left[0,\pi\right]$ for the 3D case. The symbol $\nabla$ denotes the gradient operator, and $\boldsymbol{\theta}\cdot \nabla$ represents the inner product of two vectors. For instance, we have for 2D case $\boldsymbol{\theta}\cdot\nabla=\cos\theta\partial/\partial x+\sin\theta\partial/\partial y$. If $w(\boldsymbol{x})$ is defined on a bounded domain $\Omega$ and vanishes in the exterior of the domain, the derivative can be rewritten as
\begin{equation}\label{frac_direction}
D_{\boldsymbol{\theta}}^{\alpha}w(\boldsymbol{x})=\frac{1}{\Gamma(2-\alpha)}\left(\boldsymbol{\theta}\cdot \nabla\right)^2\int_0^{d(\boldsymbol{x},\boldsymbol{\theta},\Omega)}\xi^{1-\alpha}w(\boldsymbol{x}-\xi\boldsymbol{\theta})d\xi, \quad x\in \Omega \subset \mathbb{R}^D,
\end{equation}
where the integral upper limit $d$, termed as backward distance~\cite{lischke2018fractional}, is the distance of the point $\boldsymbol{x}$ to the boundary of $\Omega$ in the direction of $-\boldsymbol{\theta}$.  The distance $d$ satisfies $\boldsymbol{x}-d(\boldsymbol{x},\boldsymbol{\theta},\Omega)\boldsymbol{\theta}\in \partial \Omega$.

Unless stated otherwise, we adopt the shifted vector Grunwald-Letnikov (GL) formula to approximate the directional fractional derivative~\cite{lischke2018fractional}:
\begin{equation}
   D_{\boldsymbol{\theta}}^{\alpha}\tilde{u}(\boldsymbol{x},t) = \frac{1}{(\Delta x)^{\alpha}}\sum_{k=1}^{\lceil \lambda d(\boldsymbol{x},\boldsymbol{\theta},\Omega)\rceil }(-1)^k \binom{\alpha}{k}\tilde{u}(\boldsymbol{x}-(k-1)\Delta x\boldsymbol{\theta},t)+O(\Delta x),
\end{equation}
where the spatial step is $\Delta x = d(\boldsymbol{x},\boldsymbol{\theta},\Omega)/\lceil \lambda d(\boldsymbol{x},\boldsymbol{\theta},\Omega)\rceil\approx 1/\lambda$. Each training point $(\boldsymbol{x},t)$ has the auxiliary points $(\boldsymbol{x}-k\Delta x\boldsymbol{\theta},t)$ that change in space for fixed time. After substituting the above formula in the fractional Laplacian definition (\ref{frac_lap}) and then applying the quadrature rule to the integral with respect to $\boldsymbol{\theta}$, we can discretize the fractional Laplacian as
\begin{multline}\label{Lx}
(-\Delta)^{\alpha/2}\tilde{u}(\boldsymbol{x},t) =C_{\alpha,D}\int_{||\boldsymbol{\theta}||_2=1}D^{\alpha}_{\boldsymbol{\theta}}\tilde{u}(\boldsymbol{x},t)d\boldsymbol{\theta}\approx \mathcal{L}^{\alpha}_{\Delta x}\tilde{u}(\boldsymbol{x},t) \\
  :=C_{\alpha,D}\int_{||\boldsymbol{\theta}||_2=1}\left\{\frac{1}{(\Delta x)^{\alpha}}\sum_{k=1}^{\lceil \lambda d(\boldsymbol{x},\boldsymbol{\theta},\Omega)\rceil }(-1)^k \binom{\alpha}{k}\tilde{u}(\boldsymbol{x}-(k-1)\Delta x\boldsymbol{\theta},t)\right\}d\boldsymbol{\theta}\\
  =\left\{ \begin{array}{c} C_{\alpha,1}\sum_{j=1}^2 \frac{1}{(\Delta x)^{\alpha}}\sum_{k=1}^{\lceil \lambda d(x,\boldsymbol{\theta}_j,\Omega)\rceil }(-1)^k \binom{\alpha}{k}\tilde{u}(x-(k-1)\Delta x\boldsymbol{\theta}_j,t) \quad (D=1) \\ C_{\alpha,2}\sum_{j=1}^{N_{\theta}} \frac{J_2\nu_j}{(\Delta x)^{\alpha}}\sum_{k=1}^{\lceil \lambda d(\boldsymbol{x},\boldsymbol{\theta}_j,\Omega)\rceil }(-1)^k \binom{\alpha}{k}\tilde{u}(\boldsymbol{x}-(k-1)\Delta x\boldsymbol{\theta}_j,t)\quad (D=2)\\
  C_{\alpha,3}\sum_{i=1}^{N_{\phi}}\sum_{j=1}^{N_{\theta}} \frac{J_3\omega_i \nu_j}{(\Delta x)^{\alpha}}\sum_{k=1}^{\lceil \lambda d(\boldsymbol{x},\boldsymbol{\theta}_{ij},\Omega)\rceil }(-1)^k \binom{\alpha}{k}\tilde{u}(\boldsymbol{x}-(k-1)\Delta x\boldsymbol{\theta}_{ij},t)\quad (D=3)
  \end{array}\right.,
\end{multline}
where $C_{\alpha,D}=\frac{\Gamma\left(\frac{1-\alpha}{2}\right)\Gamma\left(\frac{D+\alpha}{2}\right)}{2\pi^{\frac{D+1}{2}}}$ and the determinant of Jacobian matrix is $J_2=1$ for the polar-Cartesian transformation and $J_3=\sin\phi_i$ for the spherical-Cartesian transformation. The differentiation directions in the 1D, 2D and 3D cases are defined as $\{\boldsymbol{\theta}_1=1,\boldsymbol{\theta}_2=-1\}$, $\boldsymbol{\theta}_j=[\cos(\theta_j),\sin(\theta_j)]$ for $\theta_j\in(0,2\pi]$ and $\boldsymbol{\theta}_{ij}=[\sin\phi_i\cos\theta_j,\sin\phi_i\sin\theta_j,\cos\phi_i]$ for $\theta_j\in(0,2\pi],\phi_i\in[0,\pi]$, respectively. $\omega_i$ and $\nu_j$ are the Gauss-Legendre quadrature weights and $\phi_i$ and $\theta_j$ are the corresponding quadrature points. Here we take 15 quadrature points for the 2D case ($N_{\theta}=15$) and $8\times8$ quadrature points for the 3D case ($N_{\theta}=N_{\phi}=8$), which are sufficient to approximate accurately the integral with respect to the differentiation direction. Fig.~\ref{training-auxiliary} (b) provides an example of the distribution of the auxiliary points corresponding to three training points for 2D problems. We see that the collection of the auxiliary points for three distinct training points differs from each other.


\section{Numerical examples}
In this section we demonstrate the performance of fPINNs in solving forward and inverse problems of fractional ADEs. The effects of four types of numerical errors influencing the solution convergence are discussed in Section~\ref{sub_sec_forward_problem}. The solution accuracy for forward problems in different spatial dimensions is also demonstrated in the same subsection. In Section \ref{sub_sec_syn}, the solutions to inverse problems with synthetic data are shown, followed by examples with noisy data in Section~\ref{sub_sec_real}.

The fabricated solutions for time-dependent problems are given in Table~\ref{exact_solution}. The analytical forms of the space-fractional and the time-fractional derivatives of the fabricated solutions are given in \cite{dyda2012fractional} and \cite{gorenflo2014mittag}, respectively. The auxiliary function $\rho(\cdot)$ in the approximate solution (\ref{approximate_solution}) is taken as $\rho(\boldsymbol{x})=1-\|\boldsymbol{x}\|_2^2$. We consider the $L^2$ relative error of the solution predicted by fPINNs:
\begin{equation}
    \frac{\left\{\sum_k [u(\boldsymbol{x}_{test,k},t_{test,k})-\tilde{u}(\boldsymbol{x}_{test,k},t_{test,k})]^2\right\}^{\frac{1}{2}}}{\left\{\sum_k [u(\boldsymbol{x}_{test,k},t_{test,k})]^2\right\}^{\frac{1}{2}}},
\end{equation}
where $u$ and $\tilde{u}$ are fabricated and approximate solutions, respectively, and  $(\boldsymbol{x}_{test,k},t_{test,k})$ denotes the $k$-th test point. In the examples for which we compare fPINN and FDM, the test points for fPINN are selected as the finite difference grid; in the examples for which only fPINN is considered, the test points are chosen to be roughly 1000 points drawn from Sobol sequences.

\begin{table}[H]
\caption{Fabricated solutions and their fractional derivatives for time-dependent problems, where $\gamma\in (0,1]$ and $\alpha\in (1,2]$. $E_{a,b}(\cdot)$ is the Mittage-Leffler function defined by $E_{a,b}(t)=\sum_{k=0}^{\infty}\frac{t^k}{\Gamma(ak+b)}$. For $a=b=1$, it reduces to $e^{t}$.}\label{exact_solution}
\begin{center}
    \begin{tabular}{c|>{\centering\arraybackslash}p{2.5 cm}|>{\centering\arraybackslash}p{3.5 cm}|>{\centering\arraybackslash}p{7.5 cm}}
       \toprule
        & Domain &  $u(\boldsymbol{x},t)$  & $\left(\frac{\partial^{\gamma}}{\partial t^\gamma}+c(\Delta)^{\alpha/2}\right)u(\boldsymbol{x},t)$ \\
       \hline
       1D & unit interval $\{x\mid x^2\leq 1\}$ & $x(1-x^2)^{1+\frac{\alpha}{2}}e^{-t}$ &
       $\begin{array}{c}
        -t^{1-\gamma}E_{1,2-\gamma}(-t)x(1-x^2)^{1+\frac{\alpha}{2}} \\
        +c\frac{\Gamma(\alpha+3)}{6}\left[3-(3+\alpha)x^2\right]x e^{-t}
       \end{array}$
       \\
       \hline
       2D &  unit disk $\{\boldsymbol{x}\mid ||\boldsymbol{x}||_2^2\leq 1\}$ & $(1-||\boldsymbol{x}||_2^2)^{1+\frac{\alpha}{2}}e^{-t}$ &  $\begin{array}{l}
        -t^{1-\gamma}E_{1,2-\gamma}(-t)(1-||\boldsymbol{x}||_2^2)^{1+\frac{\alpha}{2}} \\
        +c 2^{\alpha}\Gamma(\frac{\alpha}{2}+2)\Gamma(\frac{2+\alpha}{2})\left[1-(1+\frac{\alpha}{2})||\boldsymbol{x}||_2^2\right] e^{-t}
       \end{array}$\\
       \hline
       3D &  unit sphere $\{\boldsymbol{x}\mid||\boldsymbol{x}||_2^2\leq 1\}$& $(1-||\boldsymbol{x}||_2^2)^{1+\frac{\alpha}{2}}e^{-t}$ & $\begin{array}{l}
        -t^{1-\gamma}E_{1,2-\gamma}(-t)(1-||\boldsymbol{x}||_2^2)^{1+\frac{\alpha}{2}} \\
        +c 2^{\alpha}\Gamma(\frac{\alpha}{2}+2)\Gamma(\frac{3+\alpha}{2})\Gamma(1.5)^{-1}\left[1-(1+\frac{\alpha}{2})||\boldsymbol{x}||_2^2\right] e^{-t}
       \end{array}$  \\
        \bottomrule
    \end{tabular}
\end{center}
\end{table}

We wrote the fPINN code in Python, and employed the Tensorflow to take advantage of its automatic differentiation capability. We also used the extended stochastic gradient descent Adam algorithm~\cite{kingma2014adam} to optimize the loss function. Unless stated otherwise, the learning rate, the number of iterations for the Adam algorithm, the number of hidden layers, and the number of neurons in each hidden layer, are fixed to be $5\times10^{-4}$, $10^5$, 4, and 20, respectively. The strategy for initializing NN parameters is given by~\cite{glorot2010understanding}.

\subsection{Forward problems} \label{sub_sec_forward_problem}
\subsubsection{Numerical convergence}\label{num_convergence}

To study the convergence of the fPINN approximation, we start with the 1D fractional Poisson problem:
\begin{equation}\label{1d-frac-Poisson}
   (-\Delta)^{\alpha/2}u(x) = f(x),\quad x\in(0,1)
\end{equation}
with the boundary conditions $u(0)=u(1)=0$. In the 1D case, the directional fractional Laplacian (\ref{frac_lap}) reduces to $(-\Delta)^{\alpha/2}:=\frac{1}{2\cos(\pi\alpha/2)}\left(D^{\alpha}_{0+}+D^{\alpha}_{1-}\right)$, where $D^{\alpha}_{0+}$ and $D^{\alpha}_{1-}$ are the left- and right-sided Riemann-Liouville fractional derivatives defined on the bounded interval $[0,1]$, respectively. They correspond to the directional fractional derivatives $D^{\alpha}_{\boldsymbol{\theta}}$ (\ref{frac_direction}) with the differentiation direction $\boldsymbol{\theta}=\pm 1$, the spatial dimension $D=1$, and the backward distances $d(x,1,[-1,1])=x$ and $d(x,-1,[-1,1])=1-x$. It should also be noted that we here assume a WB forcing term instead of a BB one in order not to deteriorate FDM solution accuracy.

The first fabricated solution we consider is $u(x)=x^3(1-x)^3$, which is smooth and hence the high-order GL formulas are valid; see Appendix B for the GL formulas of up to third-order. The corresponding forcing term is~\cite{zhao2015series}
\begin{equation}
 \begin{split}
    f(x) & = \frac{1}{2\cos(\pi\alpha/2)}\left[\frac{\Gamma(4)}{\Gamma(4-\alpha)}(x^{3-\alpha}+(1-x)^{3-\alpha})- \frac{3\Gamma(5)}{\Gamma(5-\alpha)}(x^{4-\alpha}+(1-x)^{4-\alpha})\right.\\
    & \left. \qquad \qquad \qquad+ \frac{3\Gamma(6)}{\Gamma(6-\alpha)}(x^{5-\alpha}+(1-x)^{5-\alpha})- \frac{\Gamma(7)}{\Gamma(7-\alpha)}(x^{6-\alpha}+(1-x)^{6-\alpha})\right].
  \end{split}
\end{equation}

Recalling that training and auxiliary points are generally distinct in the finite difference schemes of fPINNs, we consider $N-1$ lattice-like training points $x_j=\frac{j}{N}$ for $j=1,2,\cdots,N-1$ as well as a parameter $\lambda$ that controls the number of auxiliary points. We do not need to place training points on the boundary since the approximate solution $\tilde{u}(x)=x(1-x)u_{NN}(x;\boldsymbol{\mu})$ satisfies the boundary conditions automatically. Under this setup and considering the first-order shifted GL formula, we can write the loss function of fPINNs for problem (\ref{1d-frac-Poisson}) as
\begin{multline}
  L(\boldsymbol{\mu})=\frac{1}{N-1}\sum_{j=1}^{N-1}\left\{\frac{1}{2\cos(\alpha\pi/2)}\left[\sum_{k=0}^{\lceil\lambda d(x_j,1,[0,1])\rceil}(-1)^k\binom{\alpha}{k}\tilde{u}\left(x_j-(k-1)\frac{d(x_j,1,[0,1])}{\lceil\lambda d(x_j,1,[0,1])\rceil}\right)\right.\right.\\
  \left.\left.+ \sum_{k=0}^{\lceil\lambda d(x_j,-1,[0,1])\rceil}(-1)^k\binom{\alpha}{k}\tilde{u}\left(x_j+(k-1)\frac{d(x_j,-1,[0,1])}{\lceil\lambda d(x_j,-1,[0,1])\rceil}\right)\right]
 -f(x_j)\right\}^2.
\end{multline}
Noting that the backward distances are $d(x_j,1,[0,1])=\frac{j}{N}$ and $d(x_j,-1,[0,1])=\frac{N-j}{N}$, we rewrite the above loss function as
 \begin{multline}\label{loss-frac-PINNs}
  L(\boldsymbol{\mu})=\frac{1}{N-1}\sum_{j=1}^{N-1}\left\{\frac{1}{2\cos(\alpha\pi/2)}\left[\sum_{k=0}^{\lceil\frac{\lambda j}{N}\rceil}(-1)^k\binom{\alpha}{k}\tilde{u}\left(x_j-(k-1)\frac{\frac{j}{N}}{\lceil\frac{\lambda j}{N}\rceil}\right)\right.\right.\\
  \left.\left.+ \sum_{k=0}^{\lceil\frac{\lambda (N-j)}{N}\rceil}(-1)^k\binom{\alpha}{k}\tilde{u}\left(x_j+(k-1)\frac{\frac{N-j}{N}}{\lceil\frac{\lambda (N-j)}{N} \rceil}\right)\right]
 -f(x_j)\right\}^2.
 \end{multline}
It follows that for $\lambda=N$ the above GL finite difference scheme reduces to the scheme considered in the FDM. Actually, the FDM discretizes the equation as
\begin{multline} \label{fdm_equations}
  \frac{1}{2\cos(\alpha\pi/2)}\left[\sum_{k=0}^{j}(-1)^k\binom{\alpha}{k}\hat{u}\left(x_j-(k-1)\frac{1}{N}\right)
 + \sum_{k=0}^{N-j}(-1)^k\binom{\alpha}{k}\hat{u}\left(x_j+(k-1)\frac{1}{N}\right)\right]
 =f(x_j)
 \end{multline}
for $j=1,2,\cdots,N-1$, where $\hat{u}(\cdot)$ denotes the finite difference solution. To facilitate the comparison with the FDM, we let $\lambda=N$ for the current fabricated solution. Denoting by $\mathbf{A}_{FD}$ the finite difference matrix obtained after rearranging the coefficients in equations (\ref{fdm_equations}), we rewrite the linear system of the FDM as
\begin{equation}\label{lin_sys_fdm}
    \mathbf{A}_{FD}\left[\begin{array}{c}\hat{u}(x_1)\\ \hat{u}(x_2) \\ \vdots \\ \hat{u}(x_{N-1})\end{array}\right]=\left[\begin{array}{c}f(x_1)\\ f(x_2) \\ \vdots \\ f(x_{N-1})\end{array}\right].
\end{equation}
For $\lambda=N$, the loss function for the fPINNs is
\begin{equation}\label{loss_fun_fpinn}
    L(\boldsymbol{\mu})=MSE\left(\mathbf{A}_{FD}\left[\begin{array}{c}\tilde{u}(x_1)\\ \tilde{u}(x_2) \\ \vdots \\ \tilde{u}(x_{N-1})\end{array}\right]-\left[\begin{array}{c}f(x_1)\\ f(x_2) \\ \vdots \\ f(x_{N-1})\end{array}\right]\right),
\end{equation}
where $MSE(\mathbf{v})$ denotes the mean squared error of a vector $\mathbf{v}$. From (\ref{loss_fun_fpinn}) and (\ref{lin_sys_fdm}) we see that if arbitrary machine precision is accessible in solving linear systems and if the optimization error is negligible, then we can find an approximate solution of fPINNs such that $\tilde{u}(x_j)=\hat{u}(x_j),\forall j$ and $L(\boldsymbol{\mu})=0$. However, in practice: (1) given machine precision, say, double precision, the MSE for the FDM, i.e., $MSE(\mathbf{A}_{FD}\hat{\boldsymbol{u}}-\mathbf{f})$ can reach the squared machine precision, namely $10^{-32}$, and (2) due to optimization error, the MSE for fPINNs is much larger than that for the FDM (the lowest MSE we  obtained is $10^{-14}$). The loss function for fPINNs is high-dimensional and non-convex. For a moderate size of NN, say, a NN with 4 hidden layers (the depth of the NN is 5) and 20 neurons in each layer (the width of the NN is 20), we have more than 1000 parameters to be optimized. There are multiple local minima for the loss function, and the gradient-based optimizer almost surely gets stuck in one of the local minima~\cite{lee2016gradient}; finding global minima is a NP-hard problem~\cite{blum1989training,vsima2002training}.

\begin{figure}[H]
\centering
\includegraphics{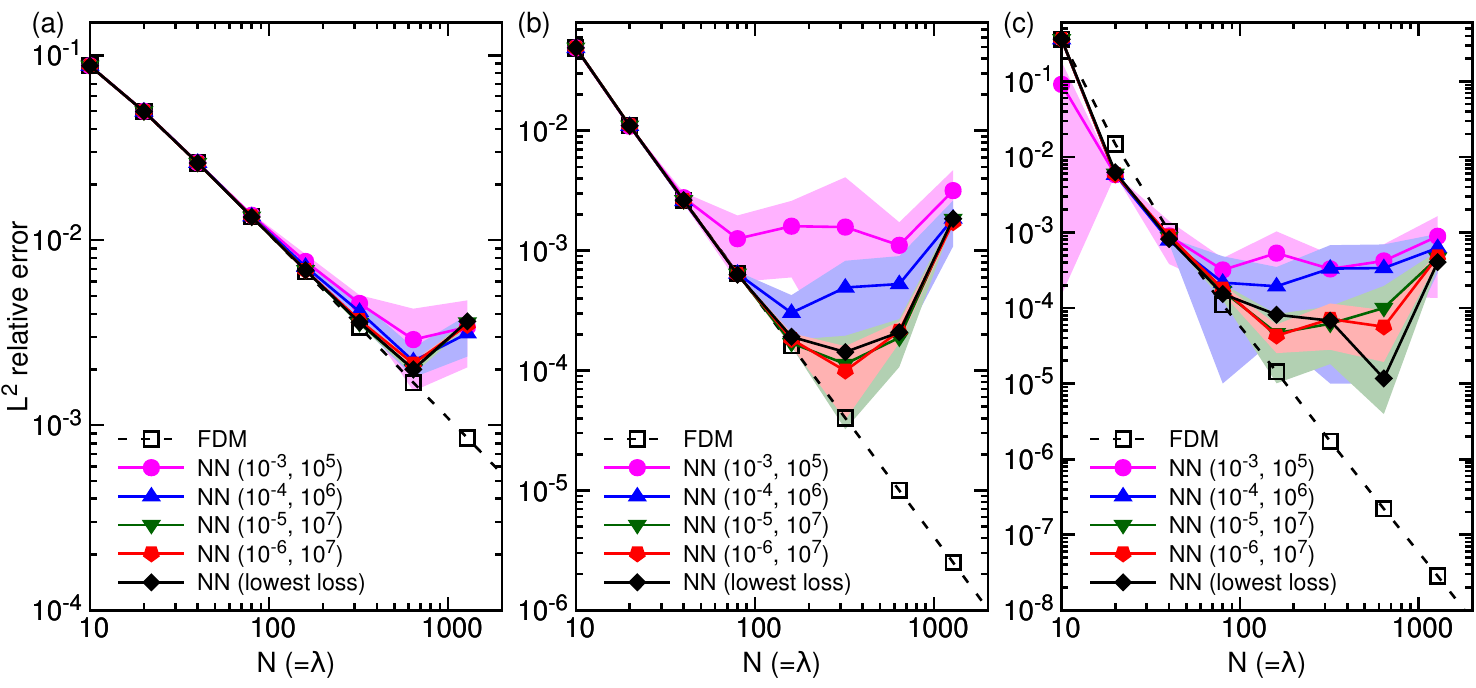}
\caption{\label{3GL-cmp}\textit{Convergence for a smooth fabricated solution}. 1D fractional Poisson equation with the fabricated solution $u(x)=x^3(1-x)^3$ and $\alpha=1.5$. Convergence for the first- (a), second- (b), and third- (c) order GL formulas versus the spatial step $\Delta x = 1/\lambda=1/N$. The dashed lines correspond to the FDM $L^2$ relative errors. The colored lines and shaded regions correspond to different learning rates (say $10^{-4}$ in the figure legend) and numbers of iterations (say $10^{6}$ in the figure legend) for the mean values and one-standard-deviation bands for the $L^2$ relative error, respectively. The solid black line denotes the errors corresponding to the lowest losses within all the learning rates cases.}
\end{figure}

In Fig.~\ref{3GL-cmp} we compare the solutions of fPINNs against FDM. We run the fPINN code 10 times, each time trying different initializations for the network weights and biases (namely $\boldsymbol{\mu}$) in order to reduce the effect of the multiple local minima. Due to the sensitivity to the initializations for the NN parameters, we plot the mean and one-standard-deviation band for the solution errors from the 10 runs, which we adopt as a new metric of convergence. To ensure the best performance of fPINNs, we consider different learning rates of the Adam optimization algorithm: $10^{-3}$, $10^{-4}$, $10^{-5}$, and $10^{-6}$. The corresponding iteration numbers are $10^5$, $10^6$, $10^7$, and $10^7$. For each learning rate, we run the code 10 times and thus we totally run the code 40 times. The solid black line in Fig.~\ref{3GL-cmp} shows the relative error corresponding to the lowest loss within the 40 runs.

We observe from Fig.~\ref{3GL-cmp} that for a fewer number of training points (a small $N$), the fPINN yields very close solution accuracy to that of the FDM, whereas for a larger number of training points, the relative error of the fPINN saturates while at the same time its uncertainty increases. From the loss function (\ref{loss-frac-PINNs}), we observe that as the number of training points $N$ increases, the number of terms in the loss function increases. The difficulty in minimizing a complex loss function is responsible for the saturation of the error for a large $N$. For fixed $N$, the complexity of the loss function is also increased when a higher order GL formula is considered. The higher the order is, the earlier the error saturates.

\begin{figure}[H]
    \centering
    \includegraphics{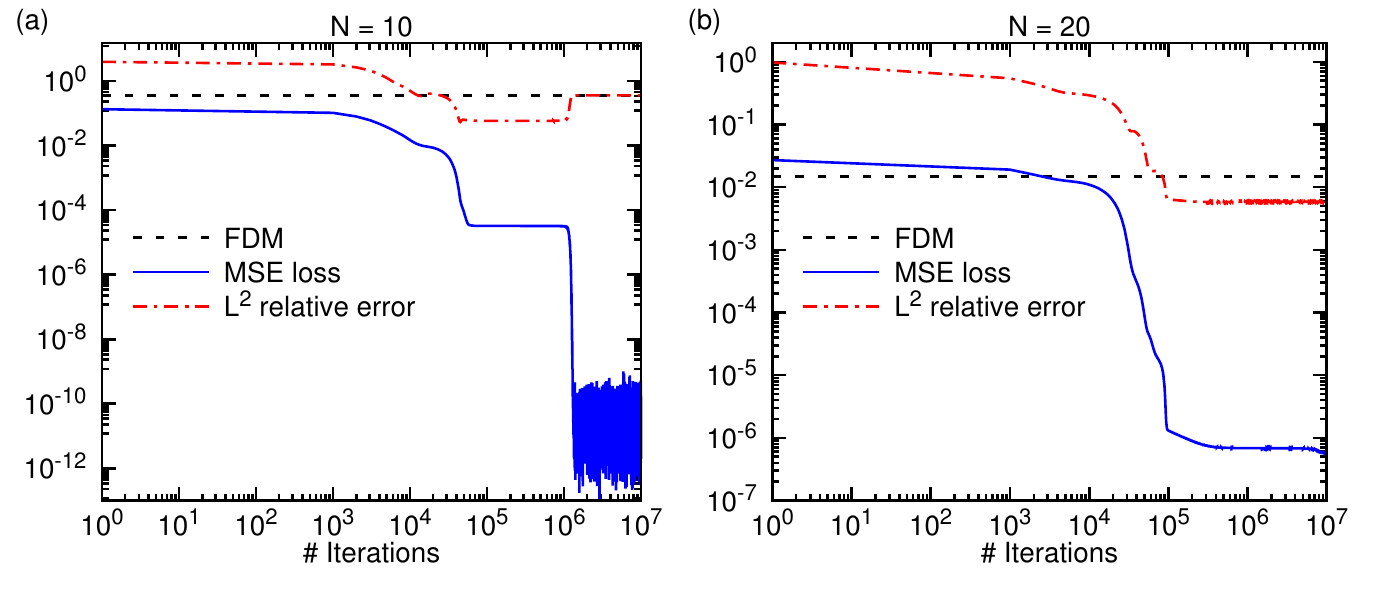}
    \caption{\textit{Bad local minimum can produce good approximate solution}. 1D fractional Poisson equation with the fabricated solution $u(x)=x^3(1-x)^3$ and $\alpha=1.5$. MSE loss and $L^2$ relative error of 10 training points (a) and 20 training points (b) versus the number of optimization iterations. Third-order GL formula is considered. The blue solid and red dashed lines are the loss and relative error, respectively. The black dashed line is the FDM error. The fPINNs code is only run once with the learning rate $10^{-6}$.}
    \label{loss-error}
\end{figure}

The solution relative error does not decrease monotonically as optimization iteration progresses. For example, Fig.~\ref{loss-error} shows the MSE loss and the relative error corresponding to the rightmost subplot of Fig.~\ref{3GL-cmp} with $N=10$ and $N=20$ and learning rate $10^{-6}$. In Fig.~\ref{loss-error} (a), between $4\times10^4$ and $10^6$ iterations, the optimizer can reach a local minimum for which the fPINN achieves lower error than the FDM. Then, after $10^6$ iterations, the optimizer can reach a better local minimum for which the relative error returns to that of the FDM. In Fig.~\ref{loss-error} (b), changing the number of training pointes from 10 to 20 increases the optimized loss from $10^{-13}$ to $10^{-7}$. More than $10^7$ iterations are needed to make the error of the fPINN return to that of the FDM.

We can explain the observations in Fig. \ref{3GL-cmp} systematically by analyzing four types of errors. These errors include the \textit{discretization error} (from the GL formula approximation), the \textit{sampling error} (from a limited number of training points), the \textit{NN approximation error} (from not sufficiently deep and wide NN), and the \textit{optimization error} (from failing to find the global minimum due to complexity of loss function). The parameters $\lambda$ and $N$ determine the discretization and sampling errors, respectively. For the current fabricated solution, we let $\lambda=N$ and thus these two errors are positively correlated. Since the fPINN and the FDM yield similar solution accuracy for small $N$, the NN approximation error is negligible. For large $N$ or $\lambda$ the optimization error dominates because the loss function becomes too complex to be optimized well.

\begin{figure}[H]
\centering
\includegraphics{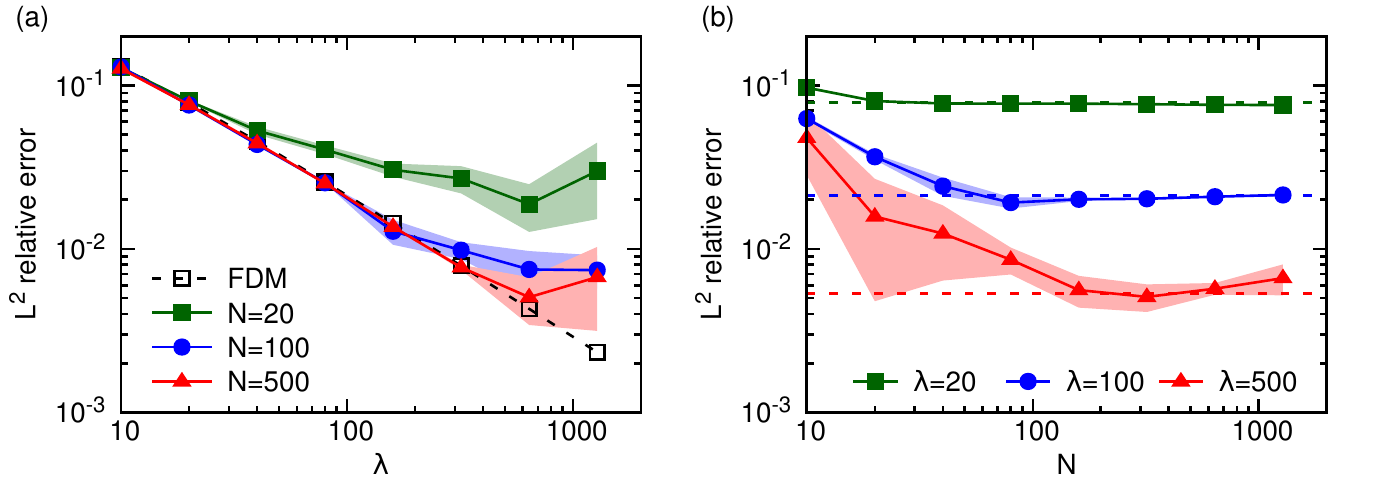}
\caption{\label{N-lambda}\textit{Convergence for a non-smooth fabricated solution}. 1D fractional Poisson equation with the fabricated solution $u(x)=x(1-x^2)^{\alpha/2}$ and $\alpha=1.5$. (a) Relative error versus the parameter $\lambda$ for fixed number of training points $N$ and (b) relative error versus $N$ for fixed $\lambda$. The dashed lines correspond to the FDM solution error. The colored lines and shaded regions correspond to mean values and one-standard-deviation bands of the fPINNs, respectively.}
\end{figure}

We will continue checking the effects of the four aforementioned errors. We consider a fabricated solution with less regularity $u(x)=x(1-x^2)^{\alpha/2}$ with the corresponding forcing term $f(x)=\Gamma(\alpha+2)x$. Unlike the previous fabricated solution case, where $\lambda=N$, we consider the lattice-like training points but take $\lambda$ different from the number of training points, $N$. We fix the sampling and the NN approximation errors by fixing $N$ and depth and width of a NN. Fig.~\ref{N-lambda} (a) shows that the discretization error dominates for small $\lambda$ until the optimization error increases sufficiently and finally dominates. In Fig.~\ref{N-lambda} (b), we fix the discretization and NN approximation errors and see that reducing the sampling error (increasing $N$) indeed makes the fPINN as accurate as the FDM. The asymptotic behavior of the error implies that there seems to be an effective number of training points, which happens when $N$ and $\lambda$ are comparable.

\begin{figure}[H]
\centering
\includegraphics{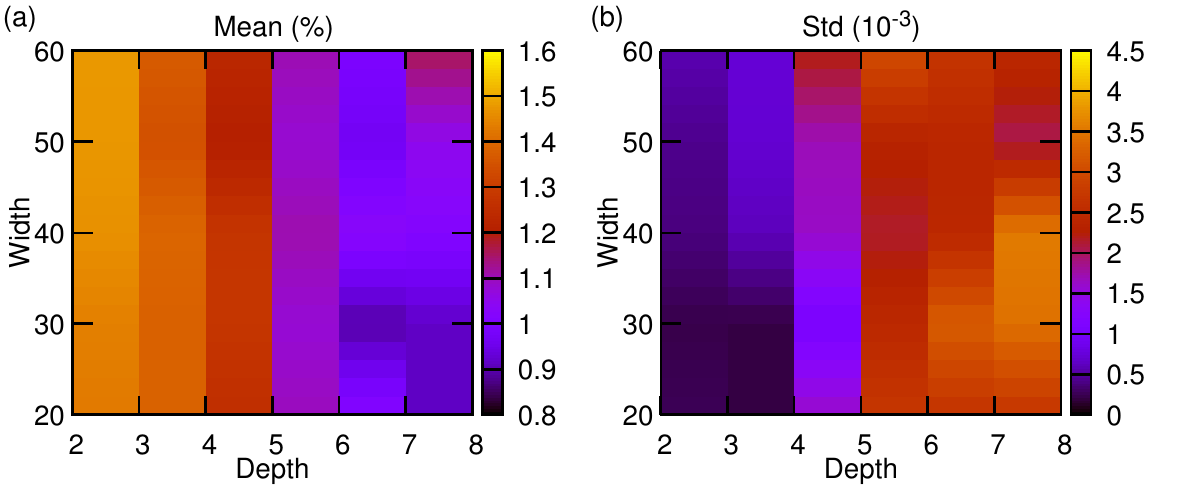}
\caption{\label{grid_search}\textit{Effect of NN architecture on the error}. 1D fractional Poisson equation with the fabricated solution $u(x)=x(1-x^2)^{\alpha/2}$ ($\alpha=1.5$), $N=100$, and $\lambda=200$. The mean (a) and the standard deviation (b) of the $L^2$ relative error versus NN depth and width. The learning rate is taken as $10^{-3}$.}
\end{figure}

\begin{figure}[H]
\centering
\includegraphics{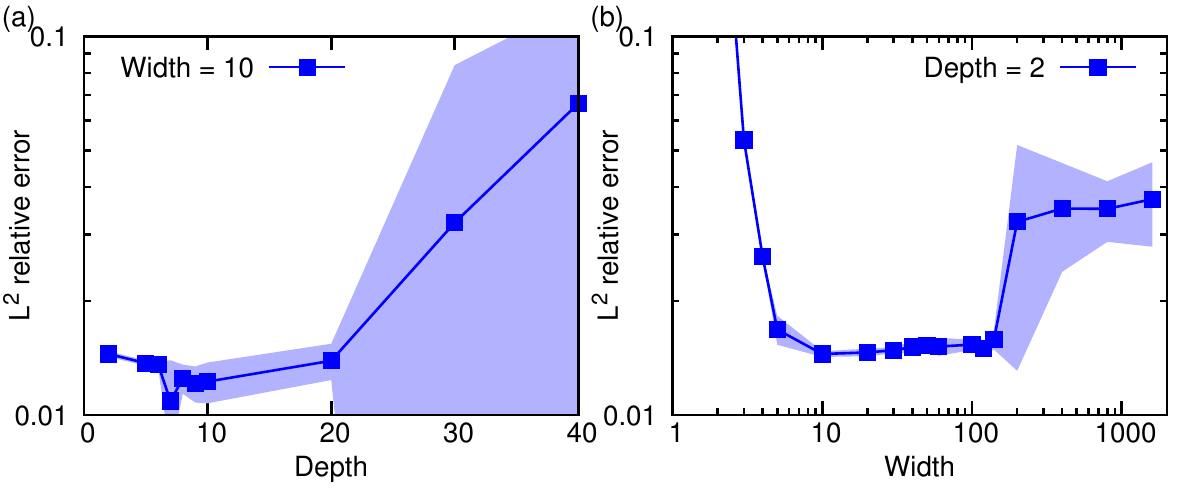}
\caption{\label{extreme-structure}\textit{Effect of extreme values of depth and width on the error}. 1D fractional Poisson equation with the fabricated solution $u(x)=x(1-x^2)^{\alpha/2}$ ($\alpha=1.5$), $N=100$, and $\lambda=200$. Errors of $u$ for a narrow NN (a) and a shallow NN (b). The learning rate is taken as $10^{-3}$.}
\end{figure}

We also study the effect of the NN \textit{approximation error} by altering the depth and width of NN. Fig.~\ref{grid_search} shows the mean and standard deviation of the relative error for different values of depth and width. We see that the depth has a larger impact on the solution accuracy than the width. Larger depth yields slightly higher accuracy but also higher uncertainty. Fig.~\ref{extreme-structure} demonstrates somewhat extreme cases for very deep or wide NNs. Large error and uncertainty are observed for large depth and width, which could be caused by overfitting (due to inadequate training points) and optimization issues. As the depth increases from 20 to 40, the error grows significantly, whereas as the width increases, the error saturates. This indicates again that the depth has larger impact on the solution accuracy than the width. From Fig.~\ref{extreme-structure} we can also observe that given other conditions, such as $N$, $\lambda$, learning rate, and iteration number, there exist an optimal depth and width for NN.

\subsubsection{Solution accuracy for time-dependent problems}

For the aforementioned time-independent problems, we introduced a new metric of convergence to represent both the mean $L^2$ relative error and standard deviation due to randomized initializations of NN parameters in training. For time-dependent problems, however, we remove the one-standard deviation bands in the following figures and simply show the relative error corresponding to the lowest loss in order to simplify the plots. However, we note that there still exist large standard deviations for a large training set (large $N$) or a large number of auxiliary points (large $\lambda$).

We consider two cases: (1) WB forcing and (2) BB forcing. In the first case, we compare the fPINN with the FDM in terms of relative error. We denote by $\lambda_t$ the parameter $\lambda$ in the temporal discretization (\ref{Lt}) and by $\lambda_x$ the parameter $\lambda$ in the spatial discretization (\ref{Lx}). The global errors for the FDM are $O(\Delta t + \Delta x)$, $O((\Delta t)^{2-\gamma}+(\Delta x)^2)$, and $O((\Delta t)^{2-\gamma}+(\Delta x))$ for the 1D space-, time-, and space-time-fractional ADEs, respectively. To facilitate the comparison, we select $\lambda_t=1/\Delta t$ and $\lambda_x=1/\Delta x$, where $\Delta t$ and $\Delta x$ are the temporal and spatial step sizes for the FDM. To ensure a constant convergence rate for the FDM, we let the temporal and spatial step sizes be related. For example, for the 1D space-time-fractional ADE, enforcing $\Delta x = (\Delta t)^{2-\gamma}$ yields the convergence rate $2-\gamma$ with respect to $\Delta t$. Lattice-like and scattered training points are both considered in the fPINNs. The lattice-like points are taken exactly the same as the finite difference gridpoints, and the scattered training points are fixed to 100 points drawn from the Sobol sequences.

\begin{figure}[H]
\centering
\subfloat[]{
\includegraphics[width=.48\textwidth]{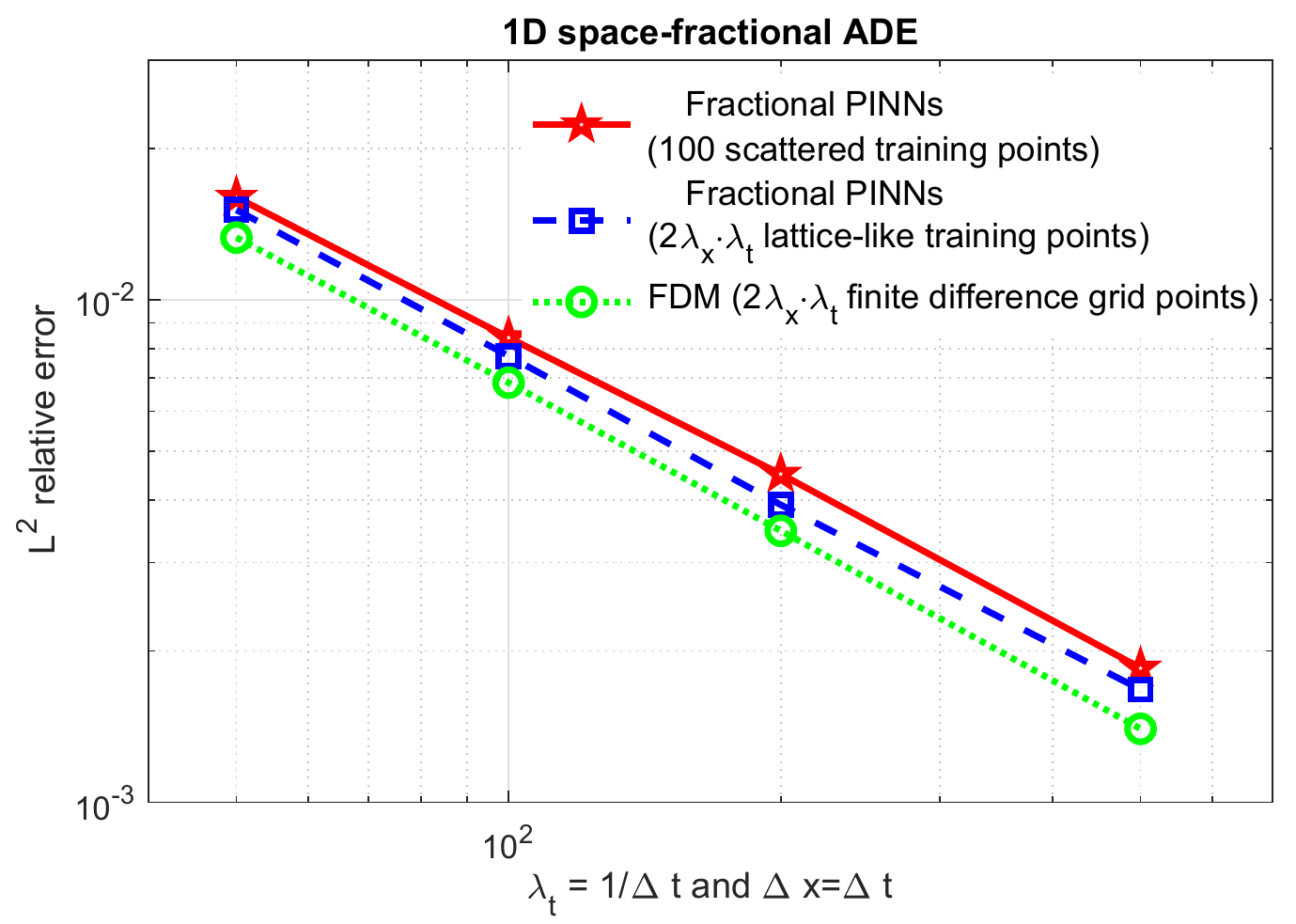}}\hfill
\subfloat[]{
\includegraphics[width=.48\textwidth]{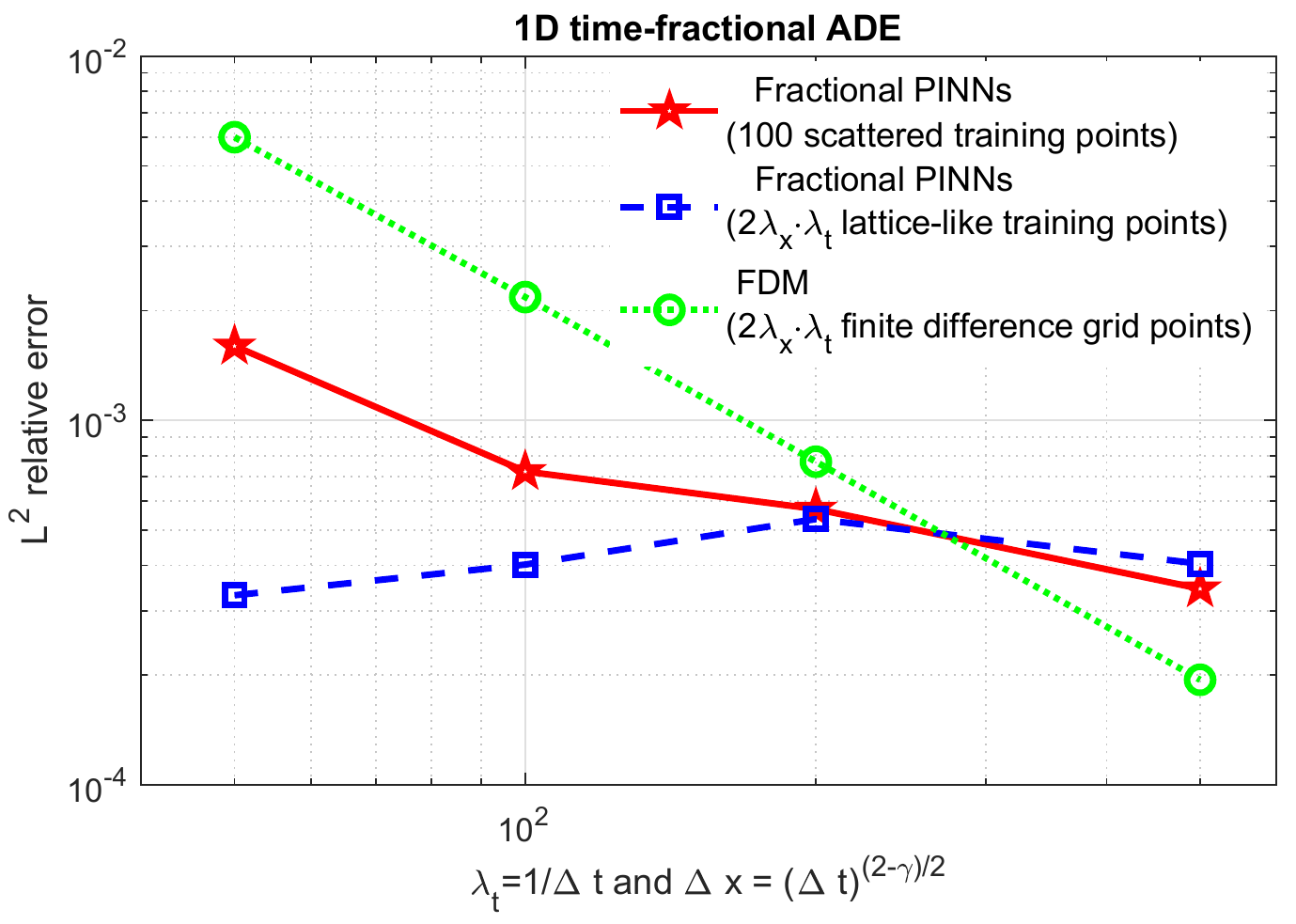}}\hfill
\subfloat[]{
\includegraphics[width=.48\textwidth]{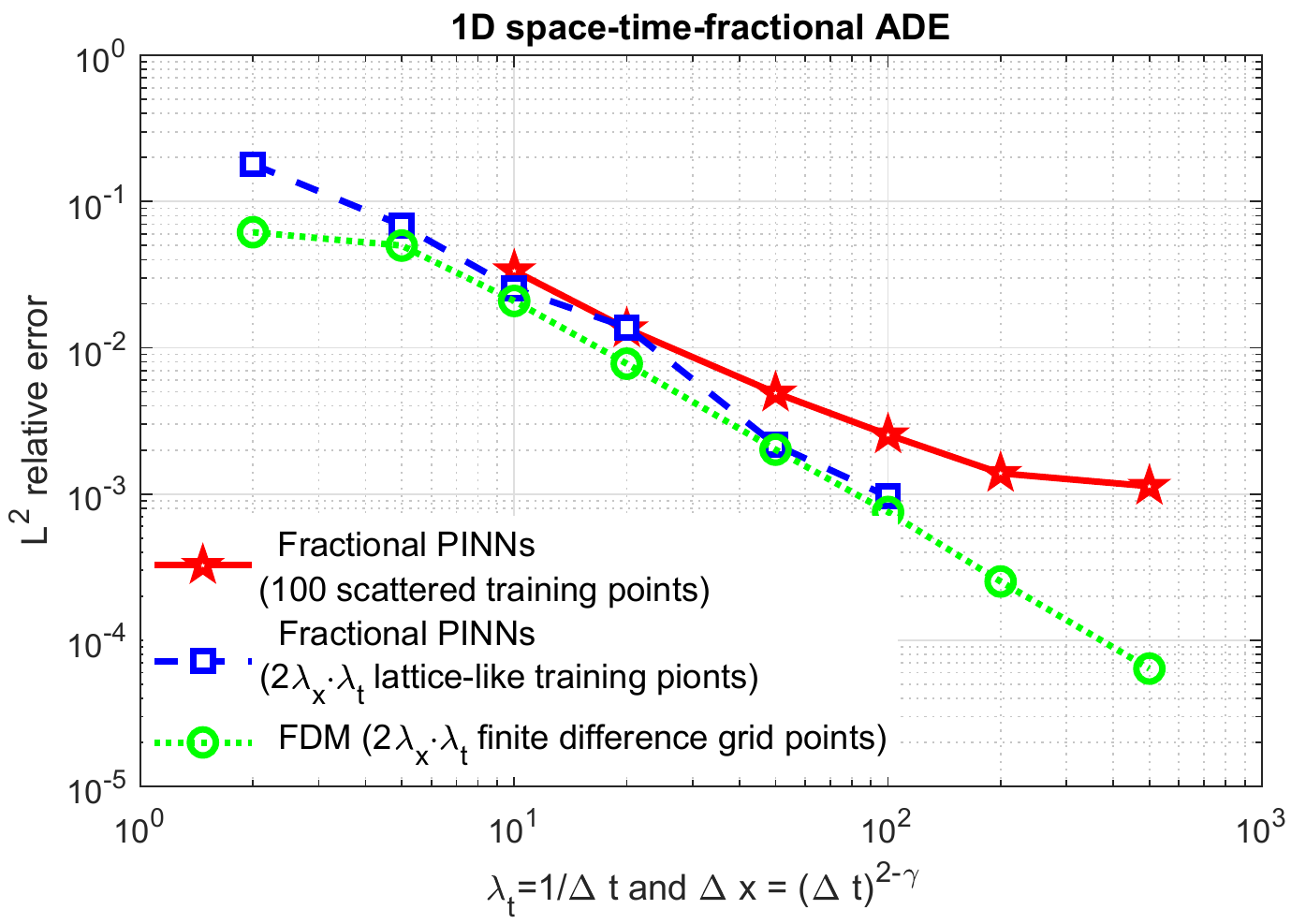}}
\caption{\label{cmp-solu-ST-frac-ADEs}\textit{Comparison of solution accuracy of the FDM and the fPINNs with scattered and lattice-like training points.} (a) 1D space-fractional ADE with $\alpha=1.5$ and $\gamma=1$. (b) 1D time-fractional ADE with $\alpha=2$ and $\gamma=0.5$. (c) 1D space-time-fractional ADE with $\alpha=1.5$ and $\gamma=0.5$. The diffusion coefficient is $c=1.0$ and the velocity is $v=0$. The fabricated solution and the corresponding forcing are given in Table~\ref{exact_solution}. In the FDM, backward difference and central difference are employed to approximate the temporal derivative in the space-fractional ADE and the spatial derivative in the time-fractional ADE, respectively.}.
\end{figure}

Fig.~\ref{cmp-solu-ST-frac-ADEs} shows the convergence of the 1D space-, time-, and space-time-fractional ADEs against the number of auxiliary points in temporal discretization. Since the fabricated solution has much lower regularity in space than in time, the spatial discretization error dominates for the three cases.
Fig.~\ref{cmp-solu-ST-frac-ADEs}~(a) shows the convergence for the 1D space-fractional ADE. The fPINN and the FDM yield close convergence rates, because the spatial discretization error dominates and the two methods have the same spatial discretization scheme. The lattice-like training points yield slightly lower accuracy than the FDM since the optimization error dominates, and they produce slightly higher accuracy than the scattered training points due to the presence of sampling error since the number of scattered training points is much lower than that of lattice-like ones.
Fig.~\ref{cmp-solu-ST-frac-ADEs}~(b) shows the convergence for the 1D time-fractional ADE. The temporal discretization errors for the fPINN and the FDM are comparable and negligible compared to the spatial numerical error. The spatial errors are different for the two methods. In the fPINN, the spatial discretization error is zero as automatic differentiation is employed to analytically derive the derivatives; the spatial numerical error mainly stems from optimization. For a small number of auxiliary points in time discretization, i.e., a small $\lambda_t$, the loss function is simple and the optimization error has little impact. Hence the fPINN outperforms the FDM. For a large $\lambda_t$ the optimization error has large impact. We can also see that the sampling error matters for the small $\lambda_t$, which explains why the lattice-like training points yield higher solution accuracy.
Fig.~\ref{cmp-solu-ST-frac-ADEs}~(c) demonstrates the convergence for the 1D space-time fractional ADE. Similar to the 1D space-fractional ADE, the spatial discretization error still dominates. Unlike the space-fractional ADE, the current ADE includes the temporal discretization and thus has more complicated loss function. The optimization error seems larger than that for the space-fractional ADE. We do not consider the lattice-like training points case for $\lambda_t>100$ for the sake of computational cost. For $\lambda_t=200$, we have $\lambda_x=1/\Delta x=1/0.005^{1.5}\approx 2828$. There are $2\cdot200\cdot2828\approx 10^6$ training points. As a result of not using mini-batch in the current training, handling such a large number of training points will be time-consuming. Additionally, we do not consider the scattered training points case for $\lambda_t<10$, since the first-order shifted GL formula does not work for a large step size if the training point is close to the boundary.

\begin{figure}[H]
\centering
\subfloat[10,20,and 50 observations]{
\includegraphics[width=.45\textwidth]{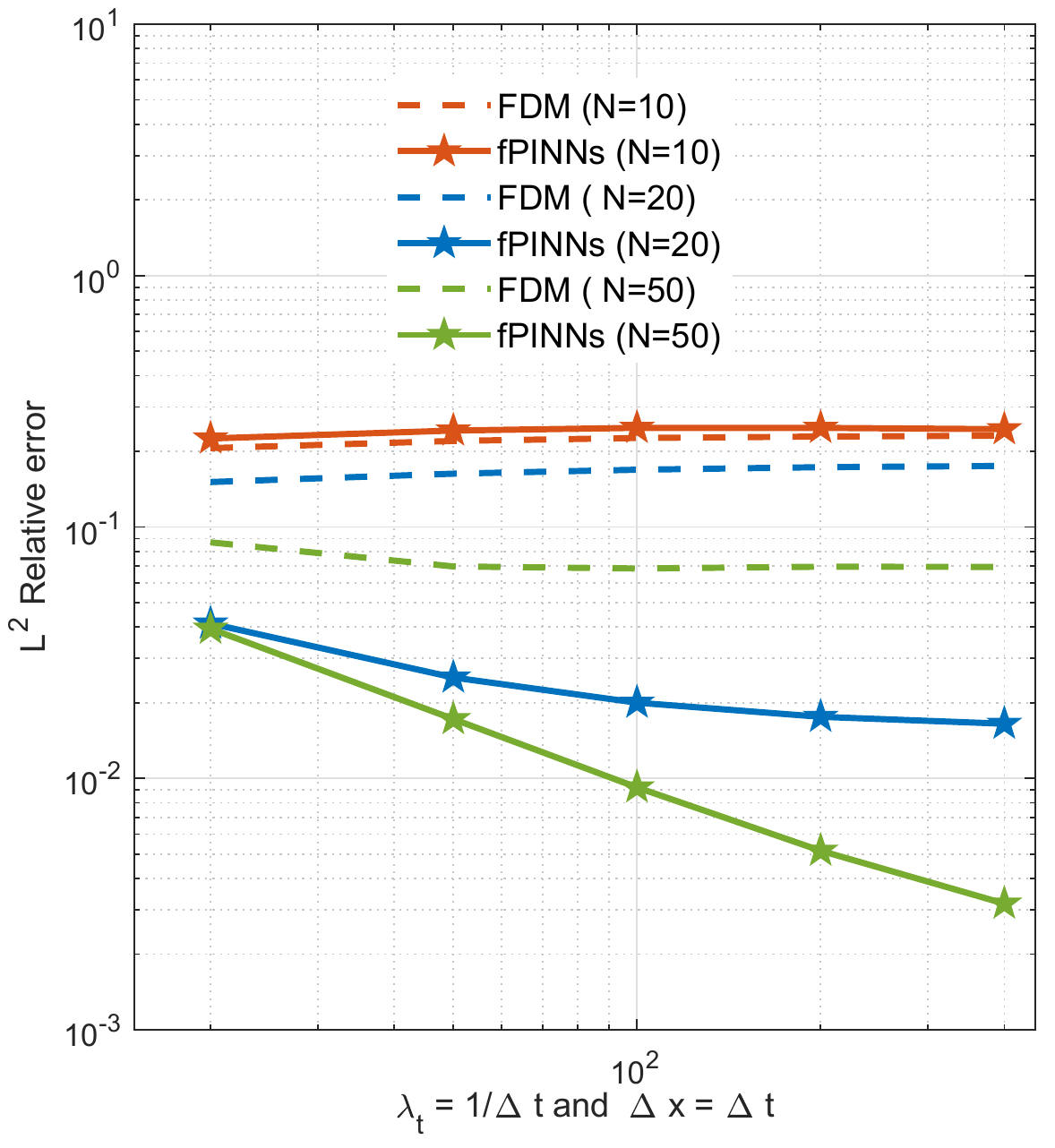}}\hfill
\subfloat[100, 400, and 1000 observations]{
\includegraphics[width=.45\textwidth]{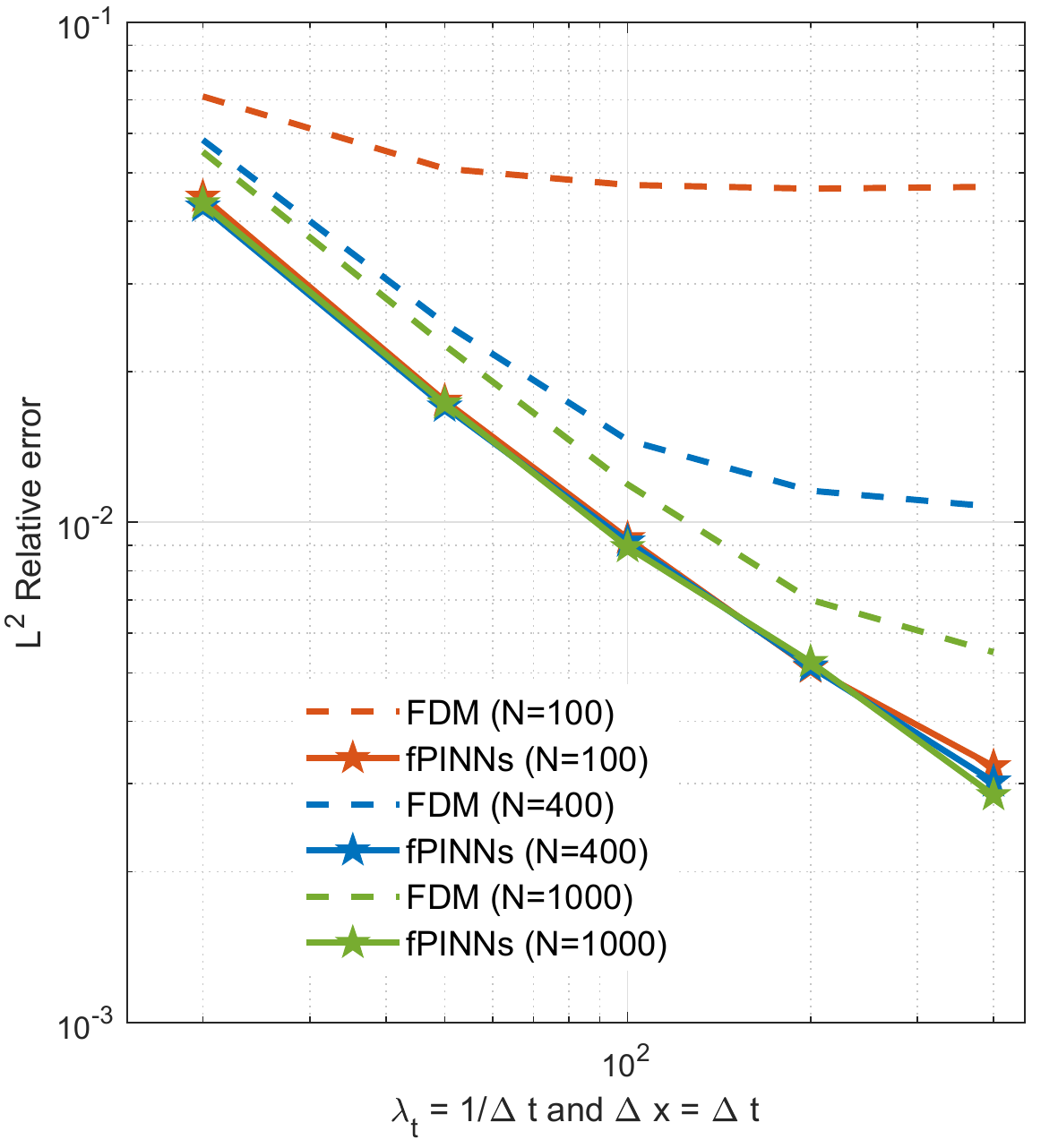}}\hfill
\caption{\label{cmp-fmd-foward}\textit{Comparison of the FDM and the fPINNs for 1D space-time-fractional problem with a black-box forcing term $f_{BB}$}. Sparse (left) and dense (right) observations for the forcing term. The fabricated solution $u(x,t)=x(1-x^2)^{1+\alpha/2}\exp(-t)$ is considered. The fractional orders are $\alpha=1.5$ and $\gamma=0.5$. The scattered training points are considered in fPINNs. The diffusion coefficient is $c=1.0$ and the velocity is $v=0.1$. A total of 100 scattered training points are taken for fPINNS, which are drawn from the Sobol sequences. In the FDM, $L_1$ scheme, first-order GL formula, and backward difference are employed to approximate the time-fractional derivative, the space-fractional derivative, and the advection term, respectively.}
\end{figure}

Although the fPINNs seem less attractive than the FDM in terms of solution accuracy for the WB forcing, they are more promising for the BB forcing. Fig.~\ref{cmp-fmd-foward} compares the two methods for the BB forcing. We observe that both methods obtain higher accuracy with the increasing number of observation points. The fPINN is superior to the FDM, especially for sparse observations ($N<50$). The low accuracy of the FDM is caused by the interpolation error for the forcing term. The fPINN directly employs the forcing values at scattered training points and no interpolation is needed.

\begin{figure}[H]
\centering
\subfloat[2D domain]{
\includegraphics[width=.2\textwidth]{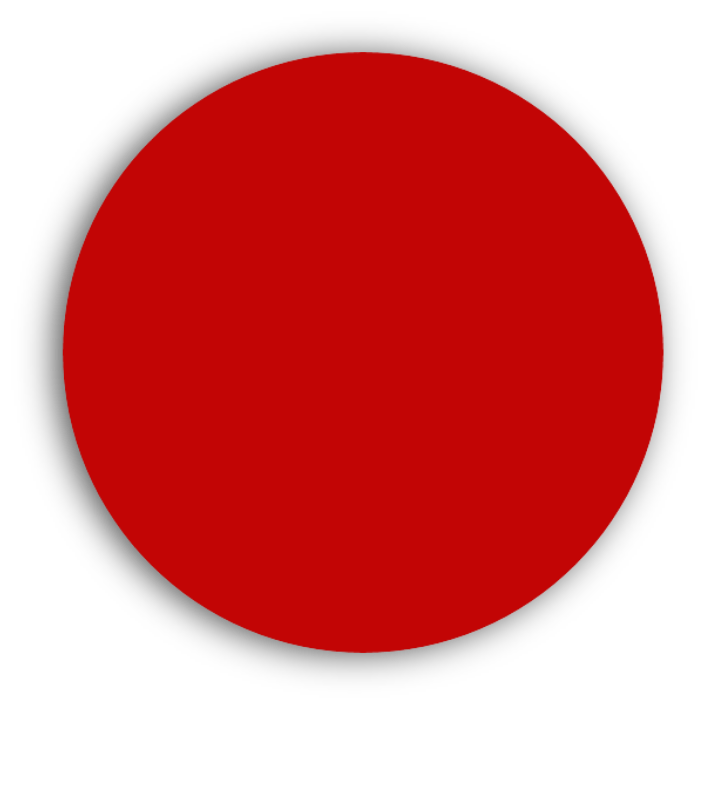}}\hfill
\subfloat[2D solution]{
\includegraphics[width=.72\textwidth]{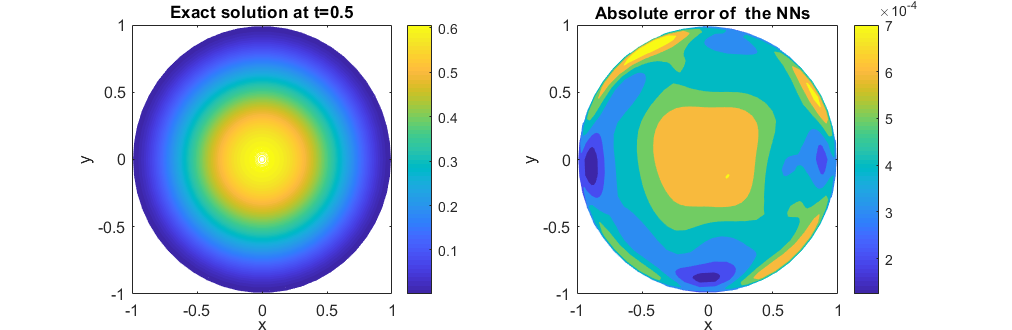}}\vfill
\subfloat[3D domain]{
\includegraphics[width=.18\textwidth]{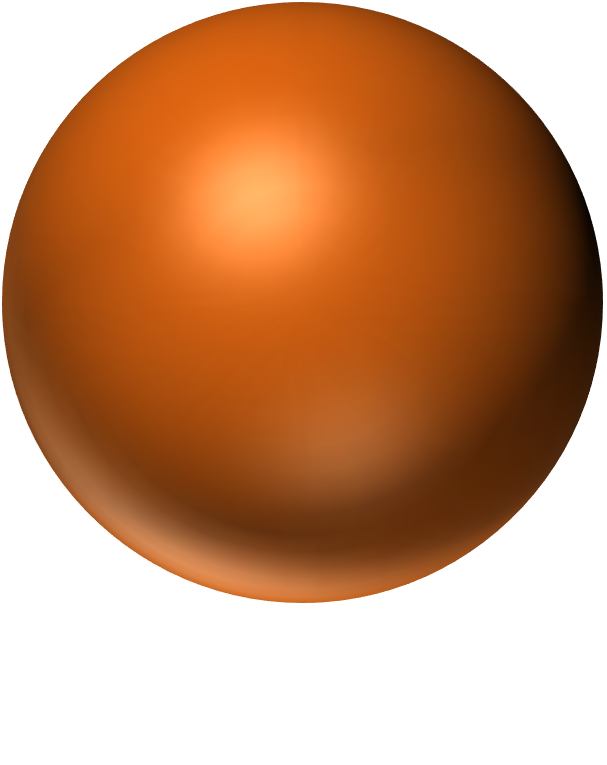}}\hfill
\subfloat[3D solution]{
\includegraphics[width=0.73\textwidth]{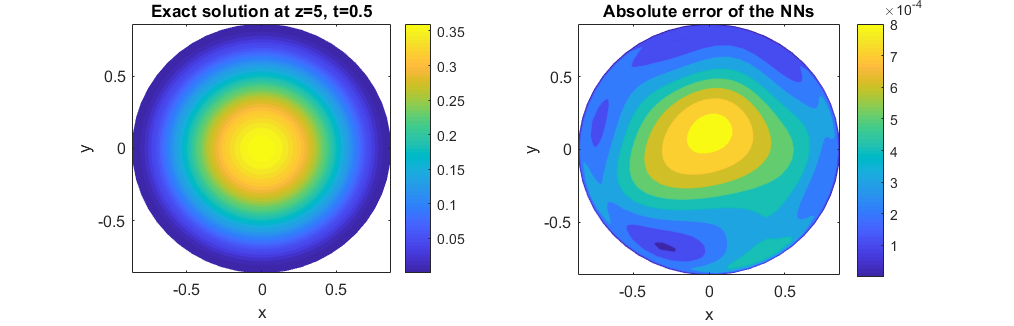}}\vfill
\caption{\textit{fPINNs accuracy in multi-dimensional simulations of space-time-fractional ADEs with the fabricated solutions given in Table \ref{exact_solution}}. (a) Unit disc computational domain. (b) exact solution (left) and the corresponding absolute error (right). (c) unit sphere computational domain. (d) exact solution (left) and the corresponding absolute error (right). The parameter setups are exactly the same as those of Table ~\ref{accuracy_forward_2d_3d}. \label{draw-2d-3d-solution}}
\end{figure}

So far very little work has been reported on the FDM for solving 2D and 3D fractional ADEs with directional fractional Laplacian. Hence, below we only show the results of fPINNs without comparison with the FDM. Table \ref{accuracy_forward_2d_3d} shows the relative errors for 2D and 3D time-dependent problems. We run the fPINN code five times and select the error corresponding to the lowest loss. We observe the relative errors from $10^{-4}$ to $10^{-3}$, which are sufficiently low. Fig. \ref{draw-2d-3d-solution} displays the contour plots of the absolute errors of the solutions in comparison with the fabricated solutions.

\begin{table}[H]
\caption{$L^2$ relative errors for 2D and 3D fractional ADEs with the fabricated solutions given in Table \ref{exact_solution}. The fractional orders are set to be (1) $\alpha=2.0,\gamma=0.5$ for the time-fractional ADE, (2) $\alpha=1.5,\gamma=1.0$ for the space-fractional ADE, and (3) $\alpha=1.5, \gamma=0.5$ for the space-time-fractional ADE. Other PDE parameters are $c=1.0$ and $v_x=v_y(=v_z)=0.1$. For 2D problems, we take $\lambda_x = 1000$ and/or $\lambda_t=400$. For 3D problems, we take $\lambda_x = 400$ and/or $\lambda_t = 200$. We take 200 and 400 scattered training points for 2D and 3D problems, respectively. All the points are drawn from the Sobol sequences. First-order GL formula is used.} \label{accuracy_forward_2d_3d}
\begin{center}
    \begin{tabular}{c|>{\centering\arraybackslash}p{4 cm}|>{\centering\arraybackslash}p{4 cm}}
       \toprule
         &  2D  & 3D \\
       \hline
       Time-fractional ADE & 3.537e-4  & 7.068e-4 \\
       Space-fractional ADE & 1.066e-3 &  2.359e-3\\
       Space-time-fractional ADE & 1.241e-3 & 2.758e-3   \\
        \bottomrule
    \end{tabular}
\end{center}
\end{table}

\subsection{Inverse problems with synthetic data} \label{sub_sec_syn}
By using the fPINNs we can solve inverse problems with almost the same coding effort as solving forward problems. We can employ the same code to solve both forward and inverse problems, since for an inverse problem code we only need to add the PDE parameters to the list of the parameters to be optimized without changing other parts of forward problem code. Increasing the dimension of the parameter space will complicate the loss function and thus make the optimization problem more difficult. Non-convexity of the loss function requires a multi-started search strategy. We run the inverse problem code 10 times with randomized initializations for both the PDE and network parameters.

In the domain $\Omega \times (0,T)$ we choose $|\Xi_1|=$ 100, 200, and 400 training points (from the Sobol sequences) for 1D, 2D, and 3D problems, respectively. In the domain $\Omega\times \{t=T\}$ we select $|\Xi_2|=$ 20, 40, and 80 additional training points (from Latin hypercube sampling) for the 1D, 2D, and 3D problems, respectively. In addition, to obtain an unconstrained optimization problem, we search a transformed parameter space $[\alpha_0,\gamma_0,c_0,\boldsymbol{v}_0]\in \mathbb{R}^{D+3}$ derived by the following transforms
\begin{equation}
    \begin{split}
        \alpha & = 0.5\tanh{(\alpha_0)}+1.5 \in (1,2),\\
        \gamma & = 0.5\tanh{(\gamma_0)}+0.5 \in (0,1),\\
        c & =\exp{(c_0)} \in (0,+\infty),\\
        \boldsymbol{v} & =\exp{(\boldsymbol{v}_0)}\in (0,+\infty)^D.
    \end{split}
\end{equation}
Noting that $f_{BB}$ and $h_{BB}$ are not identically zero, we choose the weight factors $w_1$ and $w_2$ in the loss function (\ref{frac_PINN_loss_inv}) as
\begin{equation}
    w_1=\frac{100}{\frac{1}{|\Xi_1|}\sum_{(\boldsymbol{x},t)\in\Xi_1} f^2_{BB}(\boldsymbol{x},t))},
\end{equation}
and
\begin{equation}
    w_2=\frac{1}{\frac{1}{|\Xi_2|}\sum_{(\boldsymbol{x},t)\in\Xi_2} h^2_{BB}(\boldsymbol{x},t))}.
\end{equation}
Since there are no established criteria to determine the weights $w_1$ and $w_2$, we choose them in our examples by trial and error.

\begin{figure}[H]
\centering
\subfloat[]{
\includegraphics[width=.45\textwidth]{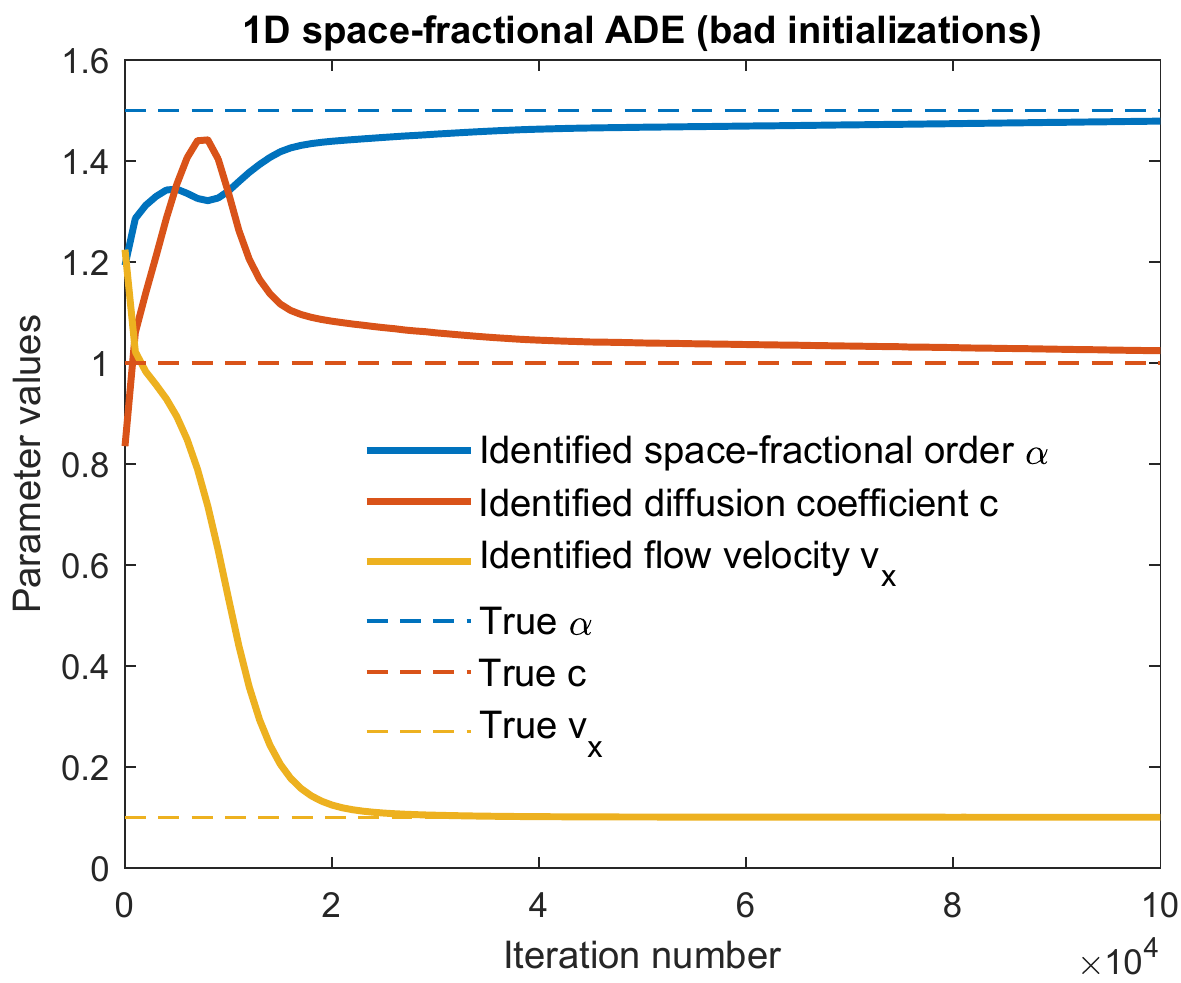}}\hfill
\subfloat[]{
\includegraphics[width=.45\textwidth]{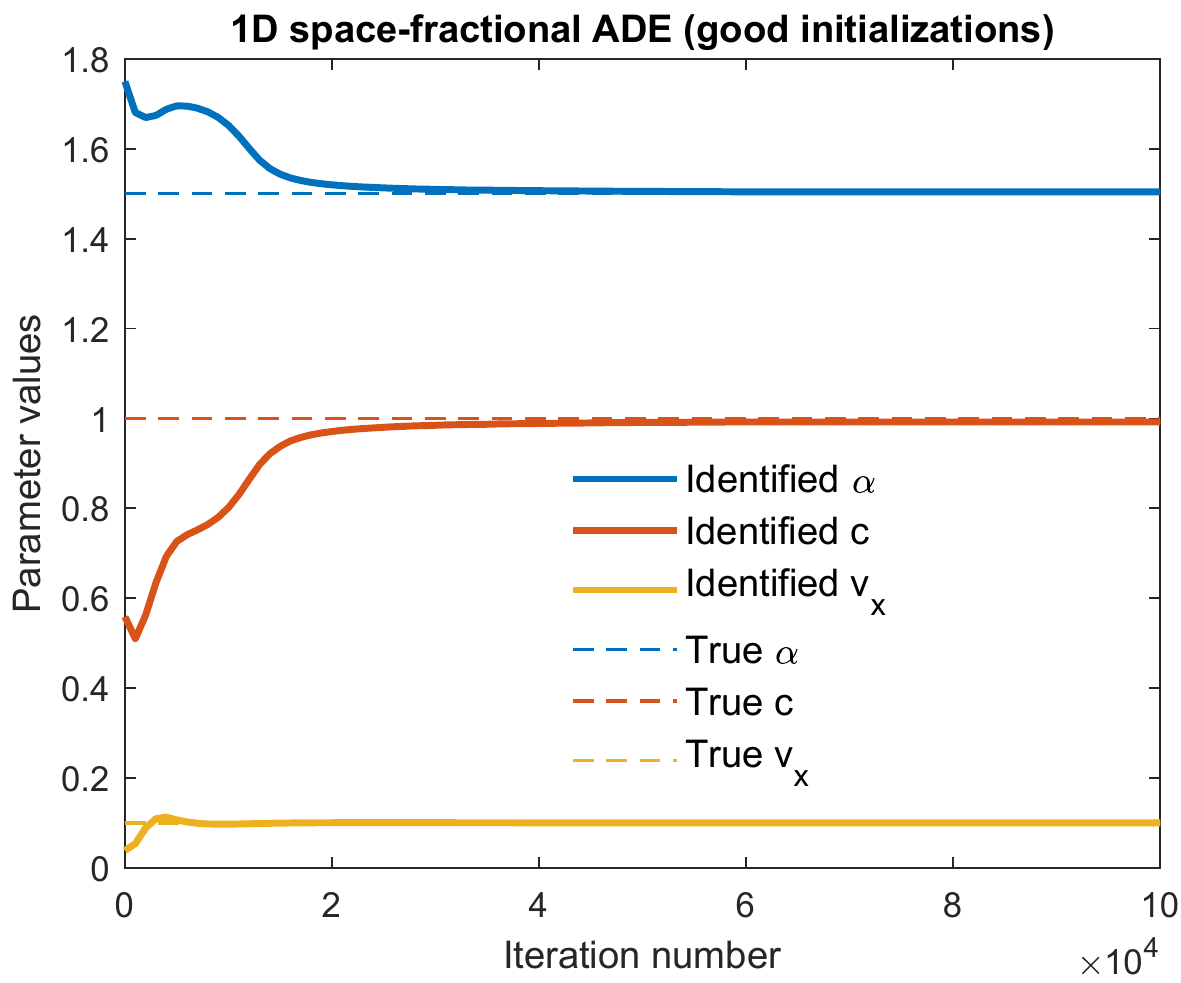}}\hfill
\caption{\label{inv-solu-1d}\textit{1D inverse problem with the fabricated solution given in Table \ref{exact_solution}}. Parameter evolution as the iteration of optimizer progresses: 1D space-fractional ADE with (a) a bad initialization and (b) a good initialization. The parameters for auxiliary points are taken as $\lambda_x=400$ and $\lambda_t=200$. The first-order GL formula is used.}
\end{figure}

\begin{figure}[H]
\centering

\includegraphics[width=.45\textwidth]{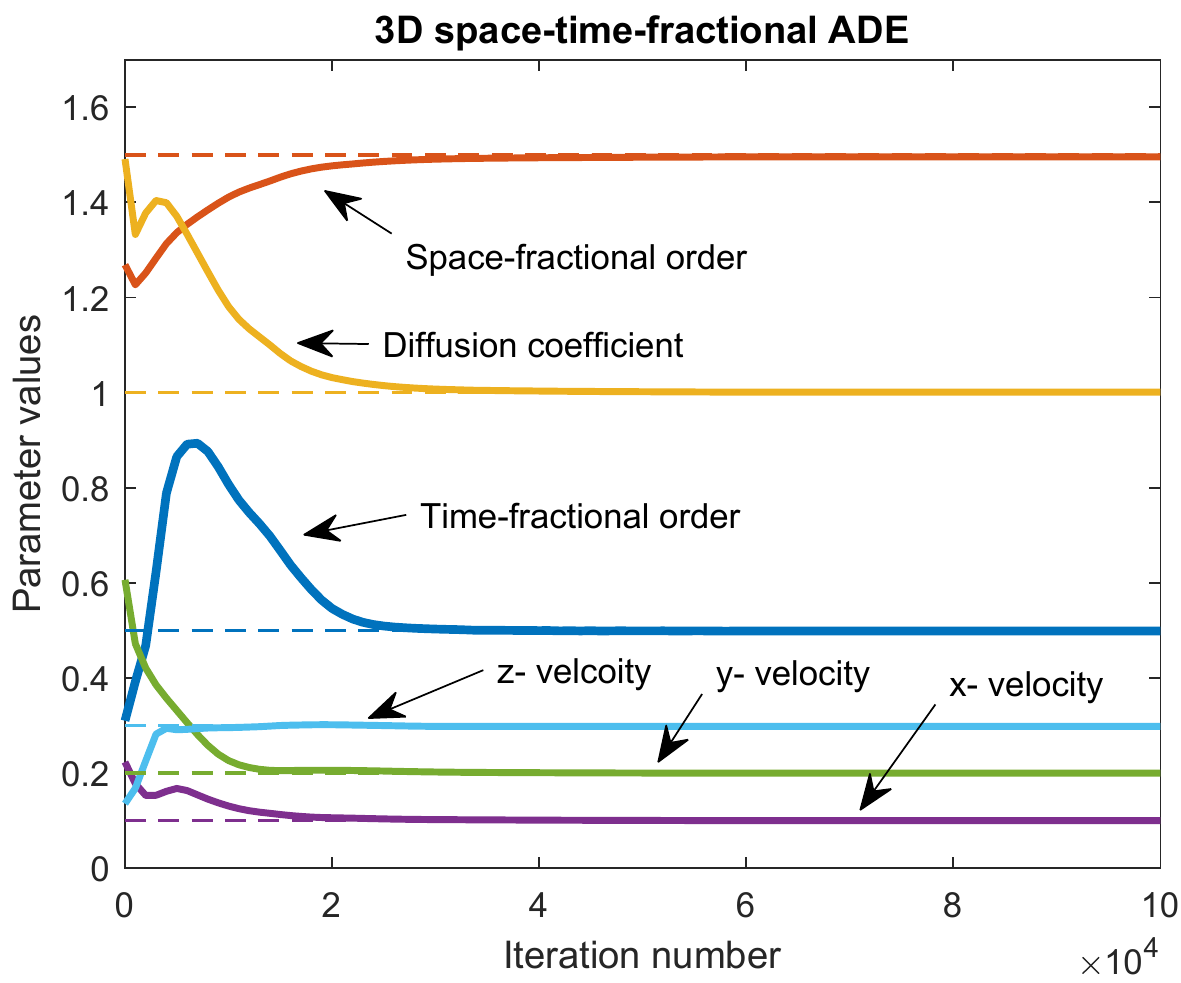}\hfill
\caption{\label{inv-3d}\textit{3D inverse problem with the fabricated solution given in Table \ref{exact_solution}}: Parameter evolution as the iteration of optimizer progresses: 3D space-time fractional ADE. The parameters for auxiliary points are taken as $\lambda_x=400$ and $\lambda_t=200$. The first-order GL formula is used.}
\end{figure}

In Fig. \ref{inv-solu-1d}, we demonstrate the trajectories of searching the parameter space for the 1D space-fractional ADE. We show two cases with a bad initialization and a good initialization. The bad and good initializations produce slow and fast convergence behaviors, respectively. By using a good initialization, we can even identify six PDE parameters of the 3D space-time-fractional ADE very well, as shown in Fig. \ref{inv-3d}.
Some strategies can be exploited to select a good initialization. For instance, we can first solve the inverse problem using small $\lambda$ and $N$ as well as a small number of optimization iterations. This preliminary result is a low-fidelity one since the solution accuracy is not high due to large errors of discretization, sampling and optimization. Then, we can use the optimized parameters from the low-fidelity problem as the good initialization.

We summarize the solutions to all the time-dependent inverse problems in Table \ref{solut-inv-all}. We can observe that both the PDE parameters and the concentration field $u$ are very well recovered.

\begin{table}[H]
\caption{Identified parameters and relative errors of the predicted concentration field $u$ for inverse problems with synthetic data. ``TF'', ``SF'', and ``STF'' are the abbreviations for time-fractional, space-fractional, and space-time fractional, respectively. We take $\lambda_x = 400$ and/or $\lambda_t=200$ for all the cases. The mean-flow velocity is denoted by $\boldsymbol{v}=[v_x,v_y,v_z]^T$ for the 3D case. }\label{solut-inv-all}
\begin{center}
    \begin{tabular}{>{\centering\arraybackslash}p{2.5 cm}|>{\centering\arraybackslash}p{4.5 cm}|>{\centering\arraybackslash}p{4.5 cm}|>{\centering\arraybackslash}p{1.5 cm}}
       \toprule
         &  True parameters  & Identified parameters & $Error_u$ \\
       \hline
       1D TF-ADE & $\gamma=0.5,c=1,v_x=0.1$  & $\gamma=0.512,c=0.999,v_x=0.102$ & 9.82e-4 \\
       2D TF-ADE & $\gamma=0.5,c=1,v_x=0.1,v_y=0.2$ &  $\gamma=0.505,c=1.000,v_x=0.0973,v_y=0.202$ & 5.430e-4\\
       3D TF-ADE & $\gamma=0.5,c=1,v_x=0.1,v_y=0.2,v_z=0.3$ & $\gamma=0.485,c=0.998,v_x=0.100,v_y=0.203,v_z=0.298$ &  8.127e-4 \\
       1D SF-ADE & $\alpha=1.5,c=1,v_x=0.1$ & $\alpha=1.476,c=1.027,v_x=0.101$ & 1.750e-3  \\
       2D SF-ADE & $\alpha=1.5,c=1,v_x=0.1,v_y=0.2$ & $\alpha=1.494,c=1.00,v_x=0.0951,v_y=0.203$ &  1.379e-3  \\
       3D SF-ADE & $\alpha=1.5,c=1,v_x=0.1,v_y=0.2,v_z=0.3$ & $\alpha=1.490,c=1.011,v_x=0.0984,v_y=0.197,v_z=0.303$ & 1.346e-3   \\
       1D STF-ADE & $\alpha=1.5,\gamma=0.5,c=1,v_x=0.1$ & $\alpha=1.501,\gamma=0.511,c=0.998,v_x=0.103$& 1.405e-3   \\
       2D STF-ADE & $\alpha=1.5,\gamma=0.5,c=1,v_x=0.1,v_y=0.2$ & $\alpha=1.5,\gamma=1.496,c=0.504,v_x=0.0945,v_y=0.199$ &  1.734e-3 \\
       3D STF-ADE & $\alpha=1.5,\gamma=0.5,c=1,v_x=0.1,v_y=0.2,v_z=0.3$ & $\alpha=1.496,\gamma=0.498,c=1.00,v_x=0.0997,v_y=0.199,v_z=0.298$ &  1.592e-3 \\
        \bottomrule
    \end{tabular}
\end{center}
\end{table}



\subsection{Forward and inverse problems using noisy data}\label{sub_sec_real}

Finally, we want to examine the effect of noisy data on system identification. To this end, for 1D problems, we add Gaussian white noise to the forcing term for the forward problem $f_{noise}(x,t) = f_{BB}(x,t) + \mathcal{N}(0,\sigma_f^2)$, where $\sigma_f=r\cdot f_{BB}(x,t)$ and to the final observation of $u$ for the inverse problem $u_{noise}(x,T)=h_{BB}(x,T)+\mathcal{N}(0,\sigma_u^2)$, where $\sigma_u=s\cdot h_{BB}(x,T)$. All other parameters are the same as those in the noise-free cases. We employ the $L^2$ regularization to smooth the noise, i.e., we add to the loss function (\ref{frac_PINN_loss_inv}) a $L^2$ norm of the vector formed by all weights and biases of NN, namely $\delta ||\boldsymbol{\mu}||_2^2$, where the regularization strength is chosen to be $\delta=10^{-4}$.

\begin{figure}[H]
\centering
\subfloat[Forward problem]{
\includegraphics[width=.4\textwidth]{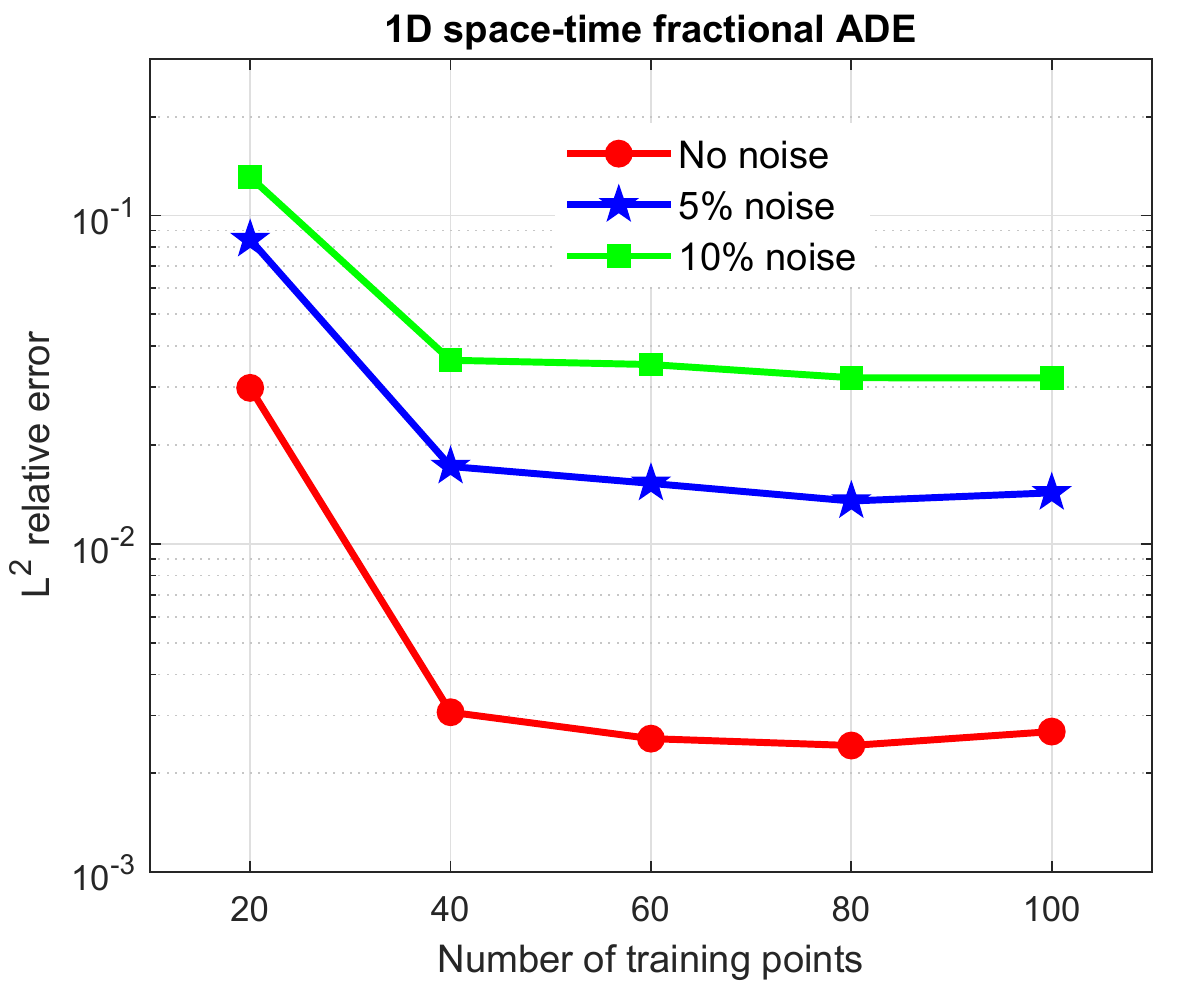}}\hfill
\subfloat[Inverse problem]{
\includegraphics[width=.55\textwidth]{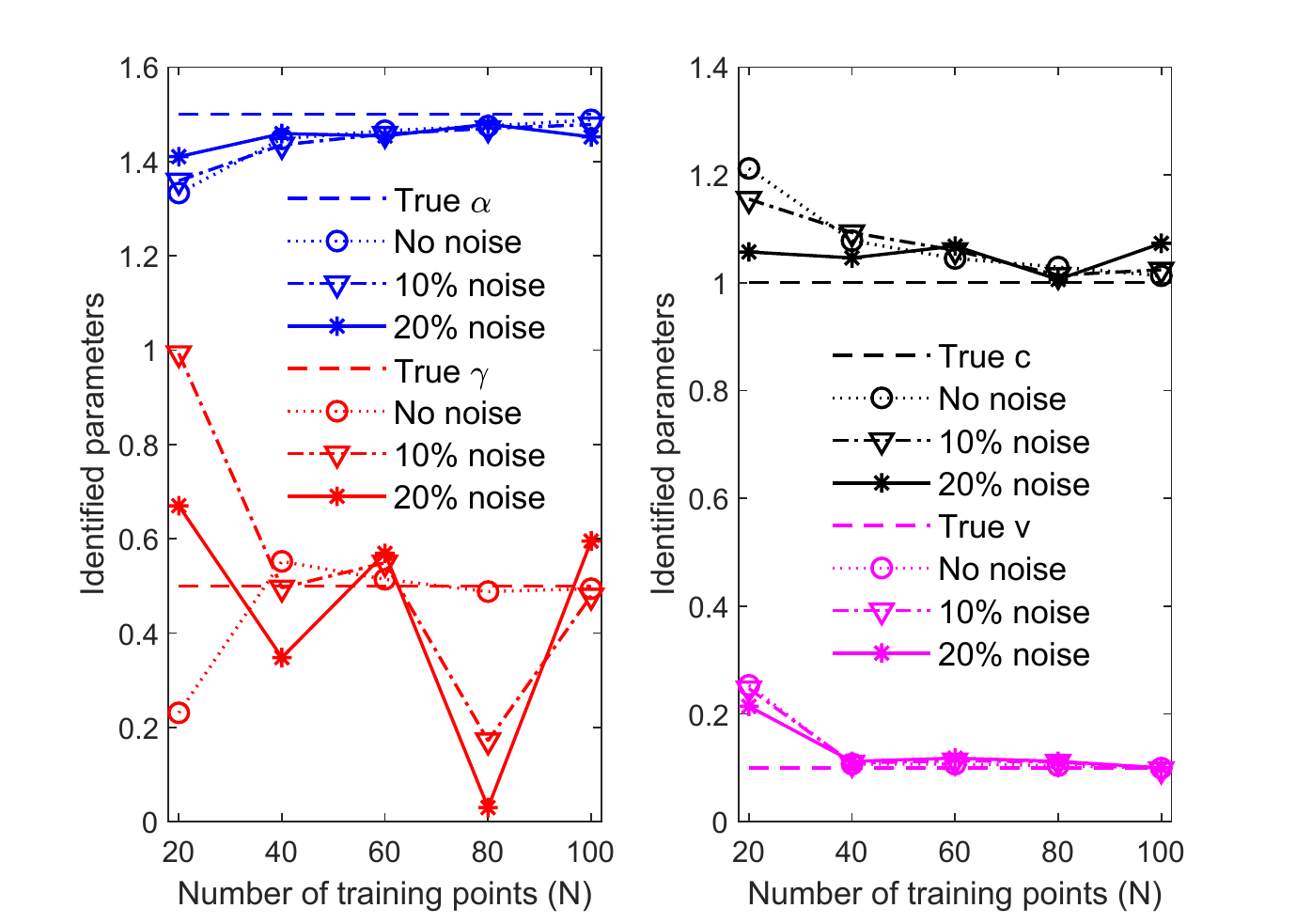}}\hfill
\caption{\label{noise-result}\textit{Influence of noise on solutions to forward and inverse problems}. (a) Forward problem of 1D space-time fractional ADE with Gaussian white noise added to the forcing term $f_{BB}$. The PDE parameters are set to be $\alpha=1.5, \gamma=0.5, c=1.0$ and $v=0.1$. (b) Inverse problem of 1D space-time fractional ADE with the Gaussian white noise added to the final observation $h_{BB}$. The $L^2$ regularization with the regularization strength $10^{-4}$ is employed to smooth the noise. NN and PDE parameters are initialized to the same values for varied noise levels and numbers of training points. For both forward and inverse problems, the parameters for auxiliary points are taken as $\lambda_x=\lambda_t=500$. }
\end{figure}

Fig.~\ref{noise-result}~(a) shows the influence of noise on solutions to the forward problem. We see that even though the forcing term is contaminated by $r=5\%$ noise, the fPINNs can still attain nearly $1\%$ relative error. Additionally, at least 40 training points or sensors are needed to recover the concentration field with roughly $1\%$ accuracy under $5\%$ noise.
Fig.~\ref{noise-result}~(b) displays the four identified parameters from noisy final observations with up to $s=20\%$ noise. The parameters exhibit different sensitivities to noise level and number of sensors. The time-fractional order $\gamma$ is most sensitive to the aforementioned two factors. To accurately identify $\gamma$ under a large amount of noise, we need to have a large number of sensors, say, at least 100 sensors for the current example. In contrast, the velocity $v$ is least sensitive to the two factors, so we can use a small number of sensors to recover the transport velocity filed if it is the only quantity to be identified.

Adding noise to $f_{BB}$ or $h_{BB}$ has different effect on the accuracy of parameter identification due to their corresponding separate contributions to the loss function. In the current example, the PDE residual has larger contribution than the final observation mismatch to the loss function and thus the noise for $f_{BB}$ will exert larger negative effect on the solution accuracy than the noise for $h_{BB}$. This explains why we can only add up to 10\% noise to $f_{BB}$ in the forward problem but we can add up to 20\% noise to $h_{BB}$ in the inverse problem.

\section{Summary and discussion}
Physics-informed neural networks (PINNs) consist of an uninformed feedforward neural network and another network induced by the physical law in the form of PDE. While previous work has focused on integer-order PDEs, here for the first time we implement a PINN that encodes fractional-order PDEs, specifically time-, space-, and space-time-fractional advection-diffusion equations (ADEs). To this end, we employ numerical differentiation formulas from fractional calculus to represent the fractional operators while we use automatic differentiation to represent the integer-order operators; we refer to this as fPINN. This framework is general and can be also applied to solution of general integral, differential, and integro-differential equations. In this mixed representation, the discretization, sampling, NN approximation, and optimization errors all affect jointly the convergence of fPINNs. Considering the 1D fractional Poisson problem, we study two different cases based on the distribution of training and finite difference discretization points: (I) $N$ lattice-like training points coinciding with $\lambda=N$ (for the computational interval of unit length) discretization points, and (II) other combinations of training and discretization points with arbitrary values of $N$ and $\lambda$. Here, we refer to discretization points as auxiliary points. In the first case, the sampling and the discretization errors are positively correlated since the training and the auxiliary points belong to the same group of points. For a few training points, the discretization (or sampling) error dominates, while the NN approximation and the optimization errors are negligible. When the number of training points increases, the optimization error dominates and the relative error of solution saturates. In the second case, the discretization and the sampling errors are generally different. For fixed sampling error, the discretization error dominates for a few auxiliary points. For fixed discretization error, the sampling error dominates for a few training points. For a large number of auxiliary or training points, the optimization error dominates. The NN approximation error depends on the NN architecture and there exist optimal depth and width for NN, but we did not pursue such systematic studies here. In addition, the optimization error stems from the non-convexity of the loss function and the optimization algorithm setup such as the learning rate and the number of iterations. Specifically, optimizing the loss function will almost surely converge to local minima~\cite{lee2016gradient} as finding the global minimum is NP-hard~\cite{blum1989training,vsima2002training}. The influences of the above errors are summarized comprehensively in Table~\ref{error-effect}.

\begin{table}[ht]
\centering
\caption{\textit{Analysis of four types of errors that influence solution convergence for 1D fractional Poisson problem.} $N$ is the number of training points; $l$ is the length of spatial domain; $\lambda$ is the parameter determining the number of auxiliary points; $\checkmark$ represents that the corresponding error dominates; $\psi\rightarrow$, $\psi\downarrow$ and $\psi\uparrow$ indicate that the value of $\psi$ is fixed, relatively small and relatively large, respectively. $\checkmark(A)$ means that the error dominates under the condition $A$. The figure numbers in the parentheses show where the conclusion comes from.}
\label{error-effect}
\begin{tabular}{l|lll}
\toprule
Error            & Sources                                                                                    & \begin{tabular}[c]{@{}l@{}}Case I: Lattice-like\\ training points and $N=l\lambda$ \end{tabular} & Case II: Other cases                           \\ \midrule
Discretization   & $\lambda$                                                                                  & \multirow{2}{*}{$\checkmark$ ($N\downarrow$, Fig.~\ref{3GL-cmp})}                                    & $\checkmark$ ($N \rightarrow$, $\lambda \downarrow$, Fig.~\ref{N-lambda}) \\ \cline{1-2} \cline{4-4}
Sampling         & $N$                                                                                        &                                                                                  & $\checkmark$ ($\lambda \rightarrow$, $N \downarrow$, Fig.~\ref{N-lambda}) \\ \hline
NN approximation & NN architecture                                                                            & \multicolumn{2}{c}{(Figs.~\ref{grid_search} and \ref{extreme-structure})}                                              \\ \hline
Optimization     & \begin{tabular}[c]{@{}l@{}}Loss function,\\ learning rate,\\ iteration number, ... \end{tabular} & \multicolumn{2}{c}{$\checkmark$ ($N\uparrow, \lambda \uparrow$, Figs.~\ref{3GL-cmp} and~\ref{loss-error})}                                                      \\ \bottomrule
\end{tabular}
\end{table}

As a data-driven approach, the fPINNs are capable of preserving high solution accuracy even when the forcing terms are black-box (BB) functions. As a mesh-free method, the fPINNs can easily handle complex-geometry computational domains in high-dimensional space. We show that  for both BB forcing and spherical computational domains the fPINNs can solve forward and inverse problems accurately. More generally, it is straightforward to consider in the fPINN representation other numerical methods instead of the Grunwald-Letnikov (GL) finite difference schemes considered in this paper. For instance, we can use the finite element method (FEM) in fPINNs. As an example, we consider the 1D steady-state fractional diffusion problem~\cite{roop2006computational}:
\begin{equation}
 \begin{split}
  \left(\frac{\partial}{\partial x}D^{\alpha-1}_{0+}-\frac{\partial}{\partial x}D^{\alpha-1}_{1-}\right)u(x)&=f(x),\quad x\in(0,1),\alpha\in(1,2),\\
  u(0)&=u(1)=0,
 \end{split}
\end{equation}
where the notation of $D^{\alpha-1}_{0+}$ and $D^{\alpha-1}_{1-}$ is the same as that in Eq.(\ref{1d-frac-Poisson}). Assuming the approximate solution to be $\tilde{u}(x)=x(1-x)u_{NN}(x;\boldsymbol{\mu})$ and integrating both hand sides of the equation with a weight function $N_j(x)$, which is a shape function corresponding to the $j$-th node, we obtain the loss function of fPINNs:
\begin{equation}
   \begin{split}
    L(\boldsymbol{\mu})&=\sum_j\left[\int_0^1\left(\frac{\partial}{\partial x}D^{\alpha-1}_{0+}-\frac{\partial}{\partial x}D^{\alpha-1}_{1-}\right) \tilde{u}(x)N_j(x)dx-\int_0^1f(x)N_j(x)dx\right]^2\\
    &=\sum_j\left[-\int_0^1\left(D^{\alpha-1}_{0+}-D^{\alpha-1}_{1-}\right) \tilde{u}(x)\frac{\partial N_j(x)}{\partial x}dx-\int_0^1f(x)N_j(x)dx\right]^2,
    \end{split}
\end{equation}
where the second summation is derived via integration by parts. The fractional derivatives of the approximate solution $\tilde{u}$ can still be approximated by using GL formulas and the integrals $\int_0^1(\cdot)dx$ can be evaluated by using Gauss quadrature. The $j$-th node corresponds to the $j$-th training point, and the nodes from the unstructured mesh correspond to scattered training points. It should be noted that, however, if the forcing term $f(x)$ is a BB function with only sparse observations, the evaluation of integral $\int_0^1f_{BB}(x)N_j(x)dx$ will be less accurate.

Despite its utility in handling sparse and noisy data and its flexibility in solving integral and/or differential equations, the fPINNs still have some limitations. First, convergence cannot be guaranteed due to the optimization error. Second, the computational cost is generally larger than that of the FDM when the two methods are both available for the same problem, as shown in Table~\ref{cost-cmp}. In future work we will investigate the form of loss function in order to avoid excessive local minima. To this end, we could change the NN architecture including the activation function type, NN width/depth, and connections between different hidden layers such as cutting and adding certain connections. We can tune these attributes of NN architecture automatically by leveraging \textit{meta-learning} techniques \cite{zoph2016neural,finn2017model}. On the other hand, we will investigate other available optimization algorithms such as in references~\cite{lei2017non,ma2017distributed} in order to expedite finding better local minima.

\begin{table}[H]
\caption{\textit{Comparison of computational cost for the FDM and the fPINNs for time-dependent problems.} In case I, the finite difference grid is $(il_x/N_x,jl_t/N_t)$ for $i=0,1,\cdots,N_x$ and $j=0,1,\cdots,N_t$ where $l_x$ and $l_t$ are the length of spatial and temporal intervals, respectively; $M$ is iteration number, $P$ denotes the average number of auxiliary points for each training point, $N_w$ represents the number of NN weights, and $N$ means the number of training points. In case I, $P\sim N_x+N_t$; in other cases, $P\sim M_{\boldsymbol{\theta}}\lambda_x+\lambda_t$, where $M_{\boldsymbol{\theta}}$ is the number of Gauss-Legendre quadrature points in (\ref{Lx}). } \label{cost-cmp}
\begin{center}
    \begin{tabular}{c|>{\centering\arraybackslash}p{6 cm}|>{\centering\arraybackslash}p{4 cm}}
      \toprule
         &  Case I: Lattice-like training points $N_x=l_x\lambda_x$,and $N_t=l_t\lambda_t$& Case II: Other cases \\
       \hline
       FDM & $O(N_{x}^3)$  & NA \\
       fPINNs & $O(MN_{x}N_tPN_w)$ & $O(MNPN_w)$ \\

        \bottomrule
    \end{tabular}
\end{center}
\end{table}

\section*{Acknowledgements}
We would like to acknowledge support from the Army Research Office (ARO)
W911NF-18-1-0301, ARO MURI W911NF-15-1-0562,  and the Department of Energy (DOE) DESC0019434, DE-SC0019453.

\bibliographystyle{unsrt}
\bibliography{ref0}

\cleardoublepage

\setcounter{equation}{0}
\renewcommand{\theequation}{A.\arabic{equation}}

\begin{appendices}\label{appendix}

\section*{Appendix A: Chain rules for integer-order and fractional derivatives}\label{appendix-A}
The $k$-th order derivative of the composite function $f(g(t))$ is given by (see Eq.(2.208) of~\cite{podlubny1999fractional})
\begin{equation}
    \frac{d^k f(g(t))}{dt^k}=k!\sum_{n=1}^k f^{(n)}(g(t))\sum\prod_{r=1}^k \frac{1}{a_r!}\left(\frac{g^{(r)}(t)}{r!}\right)^{a_r},
\end{equation}
where the second sum $\sum$ extends over all combinations of non-negative integers $a_1,a_2,\cdots,a_k$ such that $\sum_{r=1}^k ra_r=k$ and $\sum_{r=1}^k a_r = m$.

The chain rule for the Caputo fractional derivative~(\ref{caputo_der}) is as follows~\cite{diethelm2010analysis}.
\begin{equation}
 \begin{split}
   \frac{d^{\gamma}f(g(t))}{d t^{\gamma}} & =\frac{t^{-\gamma}}{\Gamma(1-\gamma)}(f(g(t)-f(g(0)))\\
   & + \sum_{k=1}^{\infty}\binom{\gamma}{k}\frac{k!t^{k-\gamma}}{\Gamma(k-\gamma+1)}\sum_{n=1}^k f^{(n)}(g(t))\sum\prod_{r=1}^k\frac{1}{a_r!}\left(\frac{g^{(r)}(t)}{r!}\right)^{a_r}.
 \end{split}
\end{equation}
It is seen that the fractional chain rule includes an infinite series $\sum_{k=1}^{\infty}$, which is computationally prohibitive when used in NN computations. For other types of fractional derivatives, particularly the fractional Laplacian, the corresponding chain rule even does not exist.

\section*{Appendix B: Grunwald-Letnikov finite difference schemes}
 Based on the stationary grid $x_j=(j-1)\Delta x$ for $j=1,2,\cdots,N$, we define the shifted GL finite difference operator for approximating the 1D fractional Laplacian
\begin{equation}
   \begin{split}
   (-\Delta)^{\alpha/2}u(x_j)\approx \delta^{\alpha}_{\Delta x,p}u(x_j) & :=(\Delta x)^{-\alpha}\sum_{k=0}^j(-1)^k\binom{\alpha}{k}u(x_j-(k-p)\Delta x)\\
   & + (\Delta x)^{-\alpha}\sum_{k=0}^{N-j}(-1)^k\binom{\alpha}{k}u(x_j+(k-p)\Delta x),
   \end{split}
\end{equation}
where we shift $p$ step size(s) to guarantee the stability of the schemes. With the notation $\beta=1-\frac{\alpha}{2}$, the first-, second-, and third-order GL formulas for approximating the 1D fractional Laplacian are as follows~\cite{zhao2015series}.
\begin{equation}
  \begin{split}
    (-\Delta)^{\alpha/2}u(x_j) & = \delta^{\alpha}_{\Delta x,1}u(x_j)+O(\Delta x),\\
    (-\Delta)^{\alpha/2}u(x_j) & = (1-\beta)\delta^{\alpha}_{\Delta x,1}u(x_j)+\beta\delta^{\alpha}_{\Delta x,0}u(x_j)+O((\Delta x)^2),\\
    (-\Delta)^{\alpha/2}u(x_j) & =  \frac{(11-6\beta)(1-\beta)}{12}\delta^{\alpha}_{\Delta x,1}u(x_j)+\frac{-6\beta^2+11\beta+1}{6}\delta^{\alpha}_{\Delta x,0}u(x_j)\\
    &+\frac{(6\beta+1)(\beta-1)}{12}\delta^{\alpha}_{\Delta x,-1}u(x_j)+O((\Delta x)^3).
  \end{split}
\end{equation}

\end{appendices}

\end{document}